\documentclass[a4paper,11pt]{article}

\usepackage[top=1.5in,right=1.0in,left=1.0in,bottom=1.0in]{geometry}
 
\usepackage{graphics,graphicx}
\usepackage{subfig}
\usepackage{float}
\usepackage{booktabs}
\usepackage{longtable}
\usepackage{supertabular}
\usepackage{tabularx}
\usepackage{multirow}
\usepackage{array}
\usepackage{tabu}
\usepackage{mathtools}
\usepackage{hyperref}
\usepackage{cancel}

\usepackage[ruled,vlined]{algorithm2e}

\usepackage{amsmath}
\usepackage{amsfonts}
\usepackage{amssymb}
\usepackage{siunitx} % to express 40km as \SI{40}{\kilometre}

\usepackage{enumerate}
\usepackage{sectsty}
\usepackage[affil-it,auth-sc]{authblk}
\usepackage{color}
\usepackage[dvipsnames]{xcolor}

\sectionfont{\normalsize}
\subsectionfont{\normalsize}

\usepackage{tikz}
\usetikzlibrary{arrows,decorations,backgrounds,shapes, snakes}
\usetikzlibrary{positioning}
\usetikzlibrary{matrix} 
\usepackage{varwidth}
\usepackage[most]{tcolorbox}
\usetikzlibrary{shapes.geometric,arrows.meta,decorations.markings}

\usepackage{biblatex}
\bibliography{bibliography}
\usepackage{graphics,graphicx}

\usepackage{tcolorbox}
\usepackage{multirow}
\usepackage{upgreek}
\usepackage{enumitem}

\usepackage{fancyhdr}            % NB: E=pagina pari, O=pagina dispari, H=intestazione, F=piede pagina, L,R,C=sinistra,destra,centro ; \leftmark = capitolo ; \rightmark = paragrafo
\fancypagestyle{plain}{%
  \fancyhead{}                                     % Pulisce il campo della testata
  \fancyfoot[CE,CO]{\thepage}       % Numero pagina in basso ed esterno in grassetto
  \fancyhead[CE,CO]{\textcolor[rgb]{0.64,0.15,0.15}{Published in \emph{European Journal of Mechanics - A/Solids} (2022) 
  \\
doi: https://doi.org/10.1016/j.euromechsol.2022.104745}}
\setlength{\headheight}{14.5pt}}
\begin{document}

%Definitions: general
\newcommand{\singlespace}{\baselineskip=12pt\lineskiplimit=0pt\lineskip=0pt}
\def\ds{\displaystyle}

\tikzstyle{every picture}+=[remember picture]

%Definitions: equations
\newcommand{\beq}{\begin{equation}}
\newcommand{\eeq}{\end{equation}}
\newcommand{\lb}{\label}
\newcommand{\ph}{\phantom}
\newcommand{\beqar}{\begin{eqnarray}}
\newcommand{\eeqar}{\end{eqnarray}}
\newcommand{\barr}{\begin{array}}
\newcommand{\earr}{\end{array}}
\newcommand{\jump}{\parallel}
\newcommand{\Ehat}{\hat{E}}
\newcommand{\That}{\hat{\bf T}}
\newcommand{\Ahat}{\hat{A}}
\newcommand{\chat}{\hat{c}}
\newcommand{\shat}{\hat{s}}
\newcommand{\khat}{\hat{k}}
\newcommand{\muhat}{\hat{\mu}}
\newcommand{\mc}{M^{\scriptscriptstyle C}}
\newcommand{\mei}{M^{\scriptscriptstyle M,EI}}
\newcommand{\mec}{M^{\scriptscriptstyle M,EC}}
\newcommand{\hbeta}{{\hat{\beta}}}
\newcommand{\rec}[2]{\left( #1 #2 \ds{\frac{1}{#1}}\right)}
\newcommand{\rep}[2]{\left( {#1}^2 #2 \ds{\frac{1}{{#1}^2}}\right)}
\newcommand{\derp}[2]{\ds{\frac {\partial #1}{\partial #2}}}
\newcommand{\derpn}[3]{\ds{\frac {\partial^{#3}#1}{\partial #2^{#3}}}}
\newcommand{\dert}[2]{\ds{\frac {d #1}{d #2}}}
\newcommand{\dertn}[3]{\ds{\frac {d^{#3} #1}{d #2^{#3}}}}
\newcommand{\ct}{\captionof{table}}
\newcommand{\cf}{\captionof{figure}}

\def\c{{\circ}}
\def\bob{{\, \underline{\overline{\otimes}} \,}}
\def\ob{{\, \underline{\otimes} \,}}
\def\scalp{\mbox{\boldmath$\, \cdot \, $}}
\def\gdp{\makebox{\raisebox{-.215ex}{$\Box$}\hspace{-.778em}$\times$}}
\def\daa{\makebox{\raisebox{-.050ex}{$-$}\hspace{-.550em}$: ~$}}
\def\mK{\mbox{${\mathcal{K}}$}}
\def\cK{\mbox{${\mathbb {K}}$}}

%Definitions: integrals
\def\Xint#1{\mathchoice
   {\XXint\displaystyle\textstyle{#1}}%
   {\XXint\textstyle\scriptstyle{#1}}%
   {\XXint\scriptstyle\scriptscriptstyle{#1}}%
   {\XXint\scriptscriptstyle\scriptscriptstyle{#1}}%
   \!\int}
\def\XXint#1#2#3{{\setbox0=\hbox{$#1{#2#3}{\int}$}
     \vcenter{\hbox{$#2#3$}}\kern-.5\wd0}}
\def\ddashint{\Xint=}
\def\fpint{\Xint=}
\def\dashint{\Xint-}
\def\cpvint{\Xint-}
\def\intl{\int\limits}
\def\cpvintl{\cpvint\limits}
\def\fpintl{\fpint\limits}
\def\ointl{\oint\limits}
\def\bA{{\bf A}}
\def\ba{{\bf a}}
\def\bB{{\bf B}}
\def\bb{{\bf b}}
\def\bc{{\bf c}}
\def\bC{{\bf C}}
\def\bD{{\bf D}}
\def\bE{{\bf E}}
\def\be{{\bf e}}
\def\bbf{{\bf f}}
\def\bF{{\bf F}}
\def\bG{{\bf G}}
\def\bg{{\bf g}}
\def\bi{{\bf i}}
\def\bI{{\bf I}}
\def\bH{{\bf H}}
\def\bK{{\bf K}}
\def\bL{{\bf L}}
\def\bM{{\bf M}}
\def\bN{{\bf N}}
\def\bn{{\bf n}}
\def\bm{{\bf m}}
\def\b0{{\bf 0}}
\def\bO{{\bf O}}
\def\bo{{\bf o}}
\def\bX{{\bf X}}
\def\bx{{\bf x}}
\def\bP{{\bf P}}
\def\bp{{\bf p}}
\def\bQ{{\bf Q}}
\def\bq{{\bf q}}
\def\bR{{\bf R}}
\def\bS{{\bf S}}
\def\bs{{\bf s}}
\def\bT{{\bf T}}
\def\bt{{\bf t}}
\def\bU{{\bf U}}
\def\bu{{\bf u}}
\def\bv{{\bf v}}
\def\bV{{\bf V}}
\def\bw{{\bf w}}
\def\bW{{\bf W}}
\def\by{{\bf y}}
\def\bz{{\bf z}}
\def\T{{\bf T}}
\def\Te{\textrm{T}}
\def\Id{{\bf I}}
\def\bxi{\mbox{\boldmath${\xi}$}}
\def\balpha{\mbox{\boldmath${\alpha}$}}
\def\bbeta{\mbox{\boldmath${\beta}$}}
\def\bepsilon{\mbox{\boldmath${\epsilon}$}}
\def\bvarepsilon{\mbox{\boldmath${\varepsilon}$}}
\def\bomega{\mbox{\boldmath${\omega}$}}
\def\bphi{\mbox{\boldmath${\phi}$}}
\def\bsigma{\mbox{\boldmath${\sigma}$}}
\def\bfeta{\mbox{\boldmath${\eta}$}}
\def\bDelta{\mbox{\boldmath${\Delta}$}}
\def\btau{\mbox{\boldmath $\tau$}}
\def\tr{{\rm tr}}
\def\dev{{\rm dev}}
\def\div{{\rm div}}
\def\Div{{\rm Div}}
\def\Grad{{\rm Grad}}
\def\grad{{\rm grad}}
\def\Lin{{\rm Lin}}
\def\Sym{{\rm Sym}}
\def\Skw{{\rm Skew}}
\def\abs{{\rm abs}}
\def\Re{{\rm Re}}
\def\Im{{\rm Im}}
\def\capB{\mbox{\boldmath${\mathsf B}$}}
\def\capC{\mbox{\boldmath${\mathsf C}$}}
\def\capD{\mbox{\boldmath${\mathsf D}$}}
\def\capE{\mbox{\boldmath${\mathsf E}$}}
\def\capG{\mbox{\boldmath${\mathsf G}$}}
\def\tcapG{\tilde{\capG}}
\def\capH{\mbox{\boldmath${\mathsf H}$}}
\def\capK{\mbox{\boldmath${\mathsf K}$}}
\def\capL{\mbox{\boldmath${\mathsf L}$}}
\def\capM{\mbox{\boldmath${\mathsf M}$}}
\def\capR{\mbox{\boldmath${\mathsf R}$}}
\def\capW{\mbox{\boldmath${\mathsf W}$}}

%imaginary unit
\def\i{\mbox{${\mathrm i}$}}
\def\mC{\mbox{\boldmath${\mathcal C}$}}
\def\mB{\mbox{${\mathcal B}$}}
\def\mE{\mbox{${\mathcal{E}}$}}
\def\mL{\mbox{${\mathcal{L}}$}}
\def\mK{\mbox{${\mathcal{K}}$}}
\def\mV{\mbox{${\mathcal{V}}$}}
\def\C{\mbox{\boldmath${\mathcal C}$}}
\def\E{\mbox{\boldmath${\mathcal E}$}}

%Definitions: journals
\def\AAM{{\it Advances in Applied Mechanics }}
\def\ACME{{\it Arch. Comput. Meth. Engng.}}
\def\ARMA{{\it Arch. Rat. Mech. Analysis}}
\def\AMR{{\it Appl. Mech. Rev.}}
\def\ASCEEM{{\it ASCE J. Eng. Mech.}}
\def\ACTA{{\it Acta Mater.}}
\def\CMAME {{\it Comput. Meth. Appl. Mech. Engrg.}}
\def\CRAS{{\it C. R. Acad. Sci. Paris}}
\def\CRM{{\it Comptes Rendus M\'ecanique}}
\def\EFM{{\it Eng. Fracture Mechanics}}
\def\EJMA{{\it Eur.~J.~Mechanics-A/Solids}}
\def\IJES{{\it Int. J. Eng. Sci.}}
\def\IJF{{\it Int. J. Fracture}}
\def\IJMS{{\it Int. J. Mech. Sci.}}
\def\IJNAMG{{\it Int. J. Numer. Anal. Meth. Geomech.}}
\def\IJP{{\it Int. J. Plasticity}}
\def\IJSS{{\it Int. J. Solids Structures}}
\def\IngA{{\it Ing. Archiv}}
\def\JAM{{\it J. Appl. Mech.}}
\def\JAP{{\it J. Appl. Phys.}}
\def\JAE{{\it J. Aerospace Eng.}}
\def\JE{{\it J. Elasticity}}
\def\JM{{\it J. de M\'ecanique}}
\def\JMPS{{\it J. Mech. Phys. Solids}}
\def\JSV{{\it J. Sound and Vibration}}
\def\MACRO{{\it Macromolecules}}
\def\MMT{{\it Mech. Mach. Th.}}
\def\MOM{{\it Mech. Materials}}
\def\MMS{{\it Math. Mech. Solids}}
\def\MMT{{\it Metall. Mater. Trans. A}}
\def\MPCPS{{\it Math. Proc. Camb. Phil. Soc.}}
\def\MSE{{\it Mater. Sci. Eng.}}
\def\NATURE{{\it Nature}}
\def\NATUREM{{\it Nature Mater.}}
\def\PHIL{{\it Phil. Trans. R. Soc.}}
\def\PMPS{{\it Proc. Math. Phys. Soc.}}
\def\PNAS{{\it Proc. Nat. Acad. Sci.}}
\def\PRE{{\it Phys. Rev. E}}
\def\PRL{{\it Phys. Rev. Letters}}
\def\PRSL{{\it Proc. R. Soc.}}
\def\RIIT{{\it Rozprawy Inzynierskie - Engineering Transactions}}
\def\ROCK{{\it Rock Mech. and Rock Eng.}}
\def\QAM{{\it Quart. Appl. Math.}}
\def\QJMAM{{\it Quart. J. Mech. Appl. Math.}}
\def\SCIENCE{{\it Science}}
\def\SCRMAT{{\it Scripta Mater.}}
\def\SM{{\it Scripta Metall.}}
\def\ZAMM{{\it Z. Angew. Math. Mech.}}
\def\ZAMP{{\it Z. Angew. Math. Phys.}}
\def\ZVDI{{\it Z. Verein. Deut. Ing.}}

\def\salto#1#2{
%\left[\mbox{\hspace{-#1em}}\left[#2\right]\mbox{\hspace{-#1em}}\right]}
[\mbox{\hspace{-#1em}}[#2]\mbox{\hspace{-#1em}}]}

\renewcommand\Affilfont{\itshape}
\setlength{\affilsep}{1em}
\renewcommand\Authsep{, }
\renewcommand\Authand{ and }
\renewcommand\Authands{ and }
\setcounter{Maxaffil}{2}

\newcommand*\diff{\mathop{}\!\mathrm{d}}

\title{%Elastic force-limiter harnessing bifurcation: \\ 
%tensile buckling and double restabilization\\ \textcolor{red}{SHOULD WE USE A MORE GENERIC TITLE?\\
Double restabilization and  design of force-displacement response\\
of the extensible elastica  with movable constraints
}
\author{P. Koutsogiannakis}
\author{D. Bigoni}
\author{F. Dal Corso\footnote{Corresponding author\\\noindent E-mail addresses:~p.koutsogiannakis@unitn.it, bigoni@ing.unitn.it, francesco.dalcorso@unitn.it}}
\affil{DICAM, University of Trento, via~Mesiano~77, I-38123 Trento, Italy}

\date{}
\maketitle
\vspace{-10mm}
\begin{center}
    \emph{Dedicated to Natasha and Sasha Movchan on the occasion of their 60th birthday}
\end{center}
\vspace{2mm}

\begin{abstract}
    A highly deformable rod, modelled as the extensible elastica, is connected to a movable clamp at one end and to a pin sliding along a frictionless curved profile at the other. Bifurcation analysis shows that axial compliance provides a stabilizing effect in compression, but unstabilizing in tension. Moreover, with varying the constraint's curvature at the origin and the axial vs bending rod's stiffness, in addition to possible buckling in tension, the structure displays   none, two, or even four bifurcation loads,  the last two associated only to the first buckling mode in compression. Therefore, the  straight configuration may lose and  recover stability one or two times, thus  evidencing single and double restabilization, 
    a feature never observed before.
     By means of the closed-form solution for the extensible elastica,  the quasi-static behaviour of the structure is analytically described under large rotations and axial strain.
    The presented  solution is exploited, together with an {\it ad hoc} developed optimization algorithm, to design  the shape of the constraint's profile necessary to obtain a  desired  force-displacement curve, so to realize a force-limiter  or a mechanical device  capable of delivering a complex force response upon application of a  continuous displacement in both positive and negative direction. 
\end{abstract}

~~~\small{Keywords: \textit{Euler buckling, tensile buckling, multistability, frictionless constraint.}}
 
\section{Introduction}

The design of structures displaying a prescribed  force-displacement response is a  challenge  in the engineering of mechanical metamaterials  \cite{ma2018origami,TAO2020108344,yu2018mechanical,zhang2021tailored}.
An important example of such a device is the  force-limiter,  used to prevent unbounded growth of a force applied to a sensitive element. In this device, when a certain threshold of  force is attained, the (incremental) stiffness  vanishes, thus allowing the displacement to grow. This occurs until a certain limit stroke is reached, at which the device finally locks. If the force-limiter is purely elastic, dissipation is prevented, so that the displacement decreases back to zero when the force is released.
Force-limiting mechanisms may be used to bound the force transmitted during an  impact, as is the case of automotive seat belt systems, where plastic deformation developing in mechanical elements prevents reuse of the device \cite{bendjellal1998combination, brumbelow2007effects, d2019forensic}. 
Another crucial application is for  vibration isolation of 
scientific and measuring equipments   \cite{alabuzhev1989vibration,ibrahim2008recent,rivin2003passive}, (e.g. optical tables \cite{vladimir2015optical}),
 or of gravitational-wave detection laboratories \cite{desalvo2007passive}, but also of civil structures to be shielded from earthquakes  
\cite{Antoniadis2018KDampingAS, kapasakalis2021kdamper,sapountzakis2017kdamper}.

Furthermore, the designed force-displacement curve can be enhanced to exhibit multistability, as in reconfigurable metamaterials for shock absorption \cite{yang2019multi}, shape programming \cite{tao20204d}, elastic energy trapping \cite{shan2015multistable}, and wave guiding \cite{jin2020guided}.
Several structural schemes have been  so far proposed for these purposes, most incorporating negative stiffness elements, such as magnetic springs for active vibration control \cite{robertson2006zero,zheng2016design,zhou2010tunable}, oblique mechanical springs \cite{carrella2007static}, extremely deformable   \cite{virgin2008vibration}, or buckling  \cite{mori2016effect,winterflood2002high} beams.

\begin{figure}[ht]
    \centering
    \includegraphics[width=150mm]{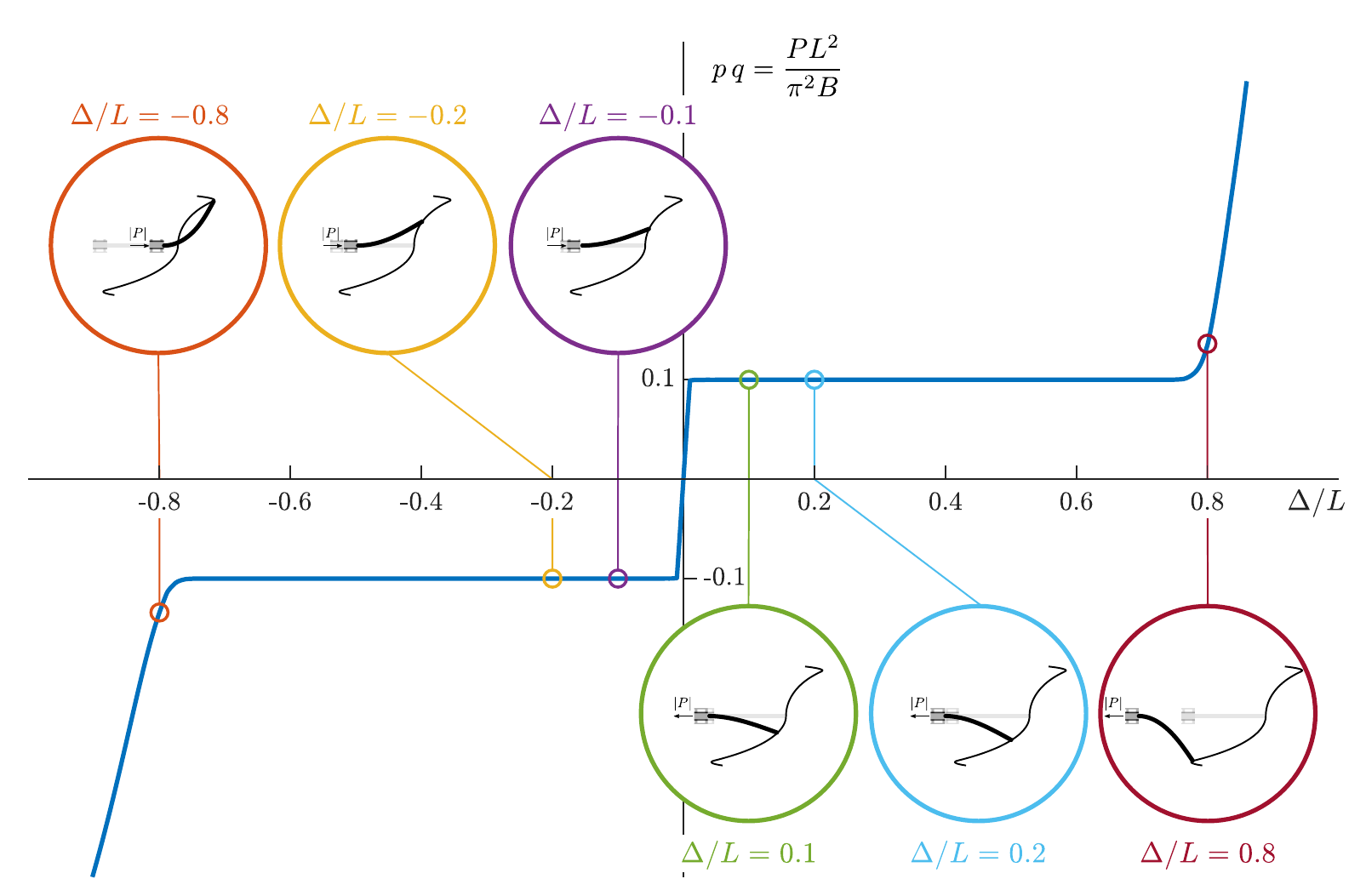}
    \caption{\label{fig:intro}  \footnotesize {Dimensionless force $P L^2/(\pi^2 B)$ vs. dimensionless displacement $\Delta/L$ for an elastic rod (flexurally and axially deformable and subject to a curved constraint), with a stiffness ratio $q=K L^2/(\pi^2 B)= 10$, optimized to realize a force limiter device  with $|P_{cr}| L^2/(\pi^2 B)=0.1$ and   locking near the end of the stroke (from  $\Delta/L=\pm0.73$). 
    Deformed configurations of the elastic rod are displayed for $\Delta/L=\pm\{0.1, 0.2, 0.8\}$, along with the optimized shape of the constraint's profile, found to have  a discontinuous curvature at the origin ($f''(0^-)=-4.186$ and $f''(0^+)=1.897$). } }
\end{figure}

Within this mechanical framework, an innovative 
structural system, based on tensile and compressive  buckling of an elastic rod, is proposed in the present article. More specifically, an  axially and flexurally deformable  rod is considered with one clamped end constrained  to move along a straight direction and the other along a curved profile with a possible discontinuous curvature. The soft rod remains straight until the axial force is sufficiently small, but it buckles (under either compression or tension) and enters in a strongly nonlinear post-critical behaviour,  when the load surpasses a critical limit, thus forcing  rod deflection. Therefore, in contrast with an inextensible rod, the soft rod deforms axially before buckling, thus allowing to display an initial linear force-displacement response. 
The considered structural system is analytically solved in its nonlinear range and an optimization algorithm is developed, allowing the design of a desired force-displacement response. 
As an example of the obtained results, 
Fig. \ref{fig:intro} reports the response of a force-limiting device designed by us and showing an excellent performance.
More in general, the force-displacement response of the proposed structure  depends only on the constraint's profile shape and on the (axial vs bending) stiffness ratio, so that negative stiffness, or sinusoidal, or triangular, or many other \lq exotic' responses can be designed (examples are
presented in Section \ref{sec:optim}).

The analysis of the proposed structure leads to another important feature, never observed before: the {\it double restabilization}  in compression.
Indeed, due to the axial compliance of the rod, it is expected that the straight configuration recovers stability after buckling, similarly to the behaviour of the \lq penetrating blade' \cite{bigoni2014instability}, and therefore displays  a restabilization in compression.
However, within a specific set of stiffness ratios and profile curvatures at the origin, the proposed structure displays  a {\it double restabilization} of the trivial (straight) configuration, where four exchanges of stability are observed. This unexpected feature is analysed in detail in Section \ref{sec:math_critical}, but a concise explanation is anticipated in Section  \ref{premessa}.

The present theoretical analysis initiates with the derivation of the nonlinear equilibrium equations for the extensible elastica (Sect. \ref{modelextela}).  Linearization of these   equations governs the determination of  bifurcation  
(Sect. \ref{sec:critical}), which is complemented by a stability analysis of the trivial configuration (by means of a dynamic approach, Sect. \ref{sec:vibr}).
The  post-critical response is evaluated   
analytically by adapting an available  closed-form solution  \cite{batista2016closed} to the present  boundary conditions (Sect. \ref{batistone}).
Finally, an optimization algorithm is developed to find the profile shape 
displaying a prescribed force-displacement curve (Sect. \ref{sec:optim}).
The solution of this problem is shown to change with varying the (axial vs bending) stiffness ratio, so that many systems can provide  the same structural response. Finally, it is shown that a broad  range of post-critical behaviours can be  achieved  (including bilinear, sinusoidal, triangular) and that more axially compliant rods facilitate the optimization. 

The results presented in this article can be used in the design of {\it passive}  mechanical devices {(i.e. not needing external control)}, delivering a designed  force-displacement response and possibly  encompassing multistable behaviour, with applications in the field of force-limiters and soft mechanisms.  
Movies of the present structure optimized to provide specific force-displacement diagrams  are available as electronic supporting material.

\subsection{A premise on (single and double) restabilization}\label{premessa}

Using linearization, a simple explanation is provided below for the destabilization and subsequent restabilization of the straight configuration of an axially and flexurally deformable elastic rod, with  a movable clamp at its left end and a pin at the  right one,
subject to an axial load $P$ that is assumed positive when tensile. 

The linearized {axial and flexural} equilibrium equations are 
\beq
         K\varepsilon(S)-P=0, ~~~
         B\,\theta''(S)-\left(1+\frac{P}{K}\right) P \,\theta(S)=F,
    \eeq
where $K$ and $B$ are the axial and bending stiffness respectively, $\varepsilon(S)$ the local axial strain, $\theta(S)$ the rotation of the rod's tangent with respect to the undeformed straight configuration, $F$ is the (unknown) reaction at the pin (orthogonal to $P$), and the prime denotes the derivative with respect to the undeformed curvilinear rod's coordinate $S$.
{
\footnote{
{For simplicity the present analysis is developed under the assumption of linear  axial behaviour. 
Although highly elastic materials such as rubber often display a nonlinear force/displacement behaviour, the linearity assumption introduced here strongly simplifies the treatment and is however adequate when the strain is not large or when the rod is realized via springs as the systems reported in Appendix \ref{ap:springs}. 
}
}
}

The boundary conditions $\theta(0)=0$, $\theta'(L)=0$, and $\int_{0}^{L}\theta(S)\,\mbox{d}S=0$ lead to the following expression for the bifurcation loads
\beq
    \pi\,\sqrt{-(1+p)\, p\,q}\, \cot\!\left(\pi \sqrt{-(1+p) \,p\, q}\right) = 1 ,
\eeq
where $p$ is the non-dimensional load and $q$ is the (axial vs bending) stiffness ratio, defined as 
\begin{equation}
\label{pera0}
    p= \frac{P}{K},\quad 
    q=\frac{K L^2}{\pi^2 B}.
\end{equation}

When the  rod is made up of a homogeneous material (and with  uniform cross-section), the stiffness ratio  reduces to  $q=\lambda^2/\pi^2\gtrsim 10^3$, where $\lambda$ is the rod's slenderness. However, values of $q$ below that range can be considered to model, via the extensible elastica, the mechanical response of architected one-dimensional structural elements.
Although the identification of such structural elements falls outside the scope of the present research,  a rough example is provided by a  rectangular wire (die) helical  spring, 
 with a high cross section aspect-ratio. Moreover, by neglecting configurational forces {
 \cite{ARMANINI201982, Bigoni2015} (which do not influence bifurcation \cite{bigoni2014instability, BOSI201583})}, 
 another example is given by  a round wire helical spring  containing a coaxial flexible piston or a spring having one end sliding along a rod with uniform cross section.\footnote{
The evaluation of the shear and axial stiffness of these structural  examples is deferred to Appendix \ref{ap:springs}.
 }

% \textcolor{red}{It is 
%  noted that  the stiffness ratio reduces to  $q=\lambda^2/\pi^2$, where $\lambda$ is the rod's slenderness, for rods with  uniform cross section. However, within a more generic setting, the extensible elastica  can be exploited to  model architected one-dimensional structures for which the stiffness ratio $q$ eludes the previous simple expression.
%  A rough example can be provided by a rod composed by two  coaxial substructures having their own axial $K_j$, and bending  $B_j$ stiffness ($j=1,2$), so that (assuming that they work in parallel) the stiffness ratio becomes $q=(K_1+K_2) L^2/(\pi^2( B_1+B_2))$ and therefore a desired $q$  value can be achieved by tuning the respective stiffnesses. An example of a structure with this property is given in Fig. \ref{fig:therod}.
% }

%  \begin{figure}
%      \centering
%      \includegraphics{figures/the_rod.eps}
%      \caption{A rod of bending stiffness $B_\text{rod}$ is inserted to a helical spring in order to enforce it's bending stiffness. The rod is allowed to slide freely within the base supporting the spring, minimizing its contribution to the axial stiffness of the structure. }
%      \label{fig:therod}
%  \end{figure}
%  }

\begin{figure}[ht]
    \centering
    \includegraphics[width=150mm]{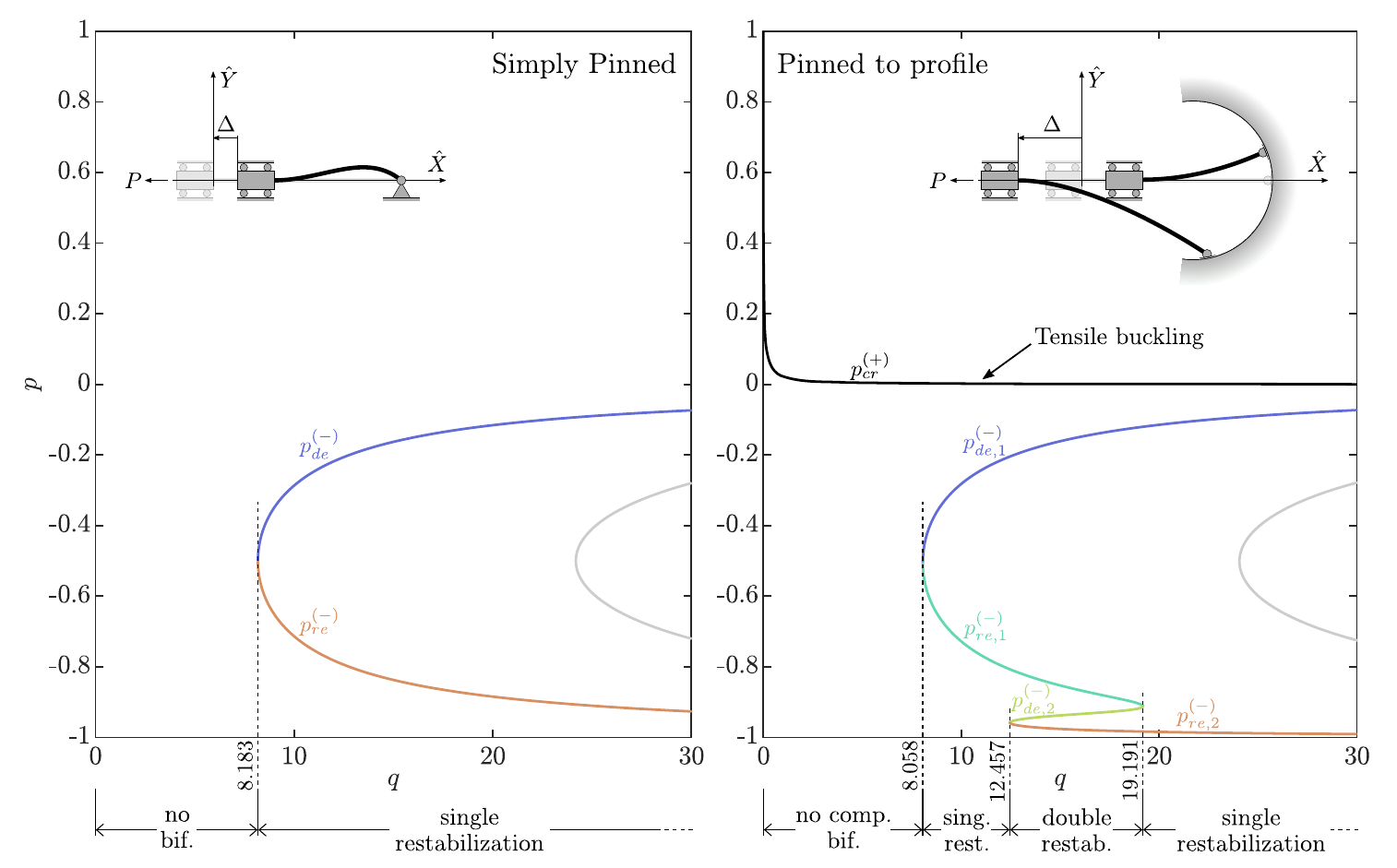}
    \caption{\label{fig:pinned} 
    \footnotesize Bifurcation load $p$ for an axially and flexurally deformable rod attached to a sliding clamp on its left end and  pinned on the right end at a fixed point (left) or at a point constrained to move on a smooth curved profile (right, with a radius of curvature equal to 1/15 of the undeformed rod's length, $f''(0)=-15$).
    The bifurcation loads  are reported at varying  axial/bending stiffness ratio $q$. The tensile buckling load $p_{cr}^{(+)}$ is reported as black line, the compressive bifurcation loads of destabilization $p_{de}^{(-)}$ and restabilization $p_{re}^{(-)}$  associated to the first mode as coloured lines, while those of higher modes as gray lines.
    Ranges of stiffness ratio $q$ for which single and double restabilization occurs are highlighted.}
\end{figure}

The bifurcation loads $p$ for an extensible rod,  subject to the above-mentioned  boundary conditions, are reported in Fig. \ref{fig:pinned} (left part) at varying the stiffness ratio  $q$. This result shows that bifurcation does not occur for $q<8.183$, while two bifurcation loads (characterized by the same instability mode) exist under compression for $q>8.183$. In the latter case, the two bifurcation loads have different mechanical interpretations: the one with smallest (greatest) absolute value is referred to  as $p_{de}^{(-)}$ $\left(p_{re}^{(-)}\right)$, because it is associated to destabilization (restabilization) of the trivial configuration at increasing compression. The structural  system therefore displays a (single) restabilization.

Anticipating some of  the results discussed in detail in section \ref{sec:math_critical}, the  presence of a curved  profile introduces  a tensile instability {\cite{zaccaria2011structures, bigoni2012effects, Bigoni2015}} at the load value $p_{cr}^{(+)}$ and a  double restabilization of the trivial configuration for $q\in (q_a,q_b)$, depending on the profile curvature. These features are shown  on  the right part of Fig. \ref{fig:pinned} for a profile with radius of curvature equal to 1/15 of the undeformed rod’s length, for which the stiffness ratio range for double restabilization is defined by $q_a=12.457$ and $q_b=19.191$.
The two bound values $q_a$ and $q_b$  will be shown to be approximately linearly increasing  with the modulus of the dimensionless (negative) profile curvature at the origin, $f''(0)$.

%%%%%%%%%%%%%%%%%%%%%%%%%%%%%%%%%%%%%%%%%%%
%%%%%%%%%%%%%%%%%%%%%%%%%%%%%%%%%%%%
It is finally  observed that the occurrence of the double restabilization requires a significant shortening of the structure, with the second (and final) restabilization displayed at a compressive deformation of more than 90\% of the undeformed length. A spring, modelled as an extensional elastica, can be \lq practically' used in order to achieve such a high compression experimentally. For example, 10 active coils with closed and ground ends of an helicoidal spring (with a wire diameter of 0.4\,mm, and  coil pitch of 10\,mm)  can achieve a shortening  of more than 95\% when compressed. 

\section{The extensible elastica with an end constrained to move along a frictionless profile}\label{modelextela}

\begin{figure}[ht!]
    \centering
    \includegraphics[width=0.8\textwidth]{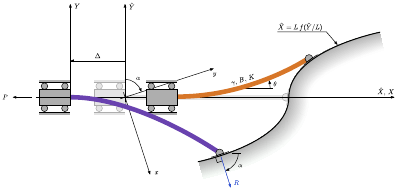}
    \caption{\label{fig:schematic} \footnotesize Scheme of the proposed structure, where an axially and flexurally deformable elastic rod is mounted on a moving clamp on its left end and slides with a pin on a curved profile on its right end. The profile is a perfectly frictionless constraint, defined by the curve $\widehat{X}=L f(\widehat{Y}/L)$. The elastic rod is in a trivial equilibrium straight configuration and may buckle  in tension and compression when a load $P$ is applied. }
\end{figure}

A soft rod is considered, connected to a sliding clamp on its left end, while the right end is constrained to move along a frictionless curved profile, Fig. \ref{fig:schematic}. The rod has a straight  undeformed configuration of length $L$ and is characterized by both axial, $K$, and flexural, $B$, stiffnesses. Therefore, it is assumed to obey  the Reissner rod model \cite{batista2016closed,reissner1972one} in which the shear stiffness is set to be infinite and the deformed configuration is described by the axial strain $\varepsilon(S)$ and the rotation $\theta(S)$ fields along the curvilinear undeformed coordinate $S\in[0,L]$.  The spatial coordinate $s$ on the deformed curve is related to the material coordinate $S$ through
\begin{equation}
    s(S) = \int_0^{S} \left[1+\varepsilon(\sigma)\right] \diff \sigma, ~~~ S \in [0, L],
\end{equation}
so that the length $\ell$ of the deformed rod is
\begin{equation}
    \ell = s(L) = \int_0^L \left[1+\varepsilon(\sigma)\right] \diff \sigma ,
\end{equation}
which differs from $L$ because of the presence of the axial strain $\varepsilon(S)$. 
By introducing a moving Cartesian reference system $X$--$Y$,  having its origin coincident with the moving clamp in the deformed configuration and with the $X$--axis aligned with the straight undeformed configuration, the coordinates of the section at a point $S$ are given by 
\begin{equation}\label{positions}
    \begin{aligned}
            & X(S; \varepsilon, \theta) = \int_0^{S} \left[1+\varepsilon(\sigma)\right] \cos\theta(\sigma) \diff \sigma,\quad
            & Y(S; \varepsilon, \theta) = \int_0^{S} \left[1+\varepsilon(\sigma)\right] \sin\theta(\sigma) \diff \sigma.
    \end{aligned}
\end{equation}

%%%%%%%%%%%%%%%%%%%%%%%%%%%%%%%%%%
%%%%%%%%%%%%%%%%%%%%%%%%%%%%%%%%%%%%%

The clamp can move only along the $ X$--axis and its  displacement from the undeformed state is measured by the distance  $\Delta$.
Introducing a fixed Cartesian reference system $\widehat{X}-\widehat{Y}$, with  origin coincident with the moving clamp in the unloaded condition and parallel to the moving $X-Y$ system, the coordinates of the rod's axis can be written as
\begin{equation}
    \begin{aligned}
            & \widehat{X}(S) = X(S; \varepsilon, \theta) - \Delta,\qquad
            & \widehat{Y}(S) = Y(S; \varepsilon, \theta) .
    \end{aligned}
\end{equation}
In this reference system, the curved profile is described by
\beq\label{eq:constraintprofile}
\widehat X\left(\widehat Y\right)=L f\left(\frac{\widehat Y}{L}\right),
\eeq
under the constraint $f(0)=1$. Moreover, the displacement  $\Delta$ (positive when opposite to the $\widehat{X}$-axis) is given by
\begin{equation}\label{eq:constraint}
    \Delta = d_X(\varepsilon, \theta) -L f\left(\dfrac{d_Y(\varepsilon, \theta)}{L}\right),
\end{equation}
where   $d_X$ is the length of the elastica projection on the $X$--axis and $d_Y$ is the vertical displacement of the pin on the right end of the rod 
\begin{equation}
\label{eq:zozza}
        d_X(\varepsilon, \theta) = \int_0^{L} \left[1+\varepsilon(\sigma)\right] \cos\theta(\sigma) \diff \sigma,\qquad
         d_Y(\varepsilon, \theta) = \int_0^{L} \left[1+\varepsilon(\sigma)\right] \sin\theta(\sigma) \diff \sigma.
\end{equation}
The moving clamp is subject to the dead load $P$, positive when opposite to the $X$ axis (namely, tensile state for the section at $S=0$), and therefore the mechanical system is conservative.
The potential energy $\mathcal V$ of the structure is the sum of the axial and flexural elastic energies and the potential energy of the dead load, 
\begin{equation}
    \mathcal V(\varepsilon, \theta) = \frac{K}{2}\int_0^{L}  \left[\varepsilon(\sigma) \right]^2\diff \sigma + \frac{B}{2}\int_0^{L}  \left[\theta'(\sigma)\right]^2 \diff \sigma - P\,\Delta,
\end{equation}
where a prime stands for differentiation with respect to the relevant (spatial) argument.
The  first variation $\delta\mathcal{V}$ of the potential energy  for perturbations in the strain $\delta \varepsilon$ and rotation $\delta \theta$ is given by
\begin{equation}
\begin{aligned}
    \delta\mathcal{V} & = \int_0^{L} K\, \varepsilon(\sigma)\, \delta\varepsilon(\sigma)\, \diff \sigma - \int_0^{L} B\, \theta''(\sigma)\, \delta\theta(\sigma)\, \diff \sigma - P\,\delta\Delta,
\end{aligned}
\end{equation} 
where
\begin{equation}
    \delta\Delta = \delta d_X - f'\left(\frac{d_Y}{L}\right)\ \delta d_Y,
\end{equation}
\begin{equation}
    \delta d_Y = \int_0^{L} [1+\varepsilon(\sigma)]\, \cos\theta(\sigma)\, \delta\theta(\sigma) \diff \sigma + \int_0^{L} \sin\theta(\sigma)\, \delta\varepsilon(\sigma)\, \diff \sigma,
\end{equation}
and
\begin{equation}
    \delta d_X = - \int_0^{L} \left[1+\varepsilon(\sigma)\right]\, \sin\theta(\sigma)\, \delta\theta(\sigma) \diff \sigma + \int_0^{L} \cos\theta(\sigma)\, \delta\varepsilon(\sigma)\, \diff \sigma.
\end{equation}

The vanishing of the first variation $\delta\mathcal{V}$ for compatible perturbations $\delta \varepsilon$ and  $\delta \theta$ provides the following coupled system of nonlinear equilibrium equations 
\begin{equation} \label{eq:governing}
    \begin{aligned}
        & B \theta''(S) - P\,\left[1+\varepsilon(S)\right]\,\left[ f'\left(\frac{d_Y}{L}\right) \,\cos\theta(S) + \sin\theta(S) \right] = 0, \\
        & K \varepsilon(S) + P\,\left[ f'\left(\frac{d_Y}{L}\right) \,\sin\theta(S) - \cos\theta(S) \right] = 0,
    \end{aligned}
\end{equation}
complemented by the algebraic constraint, Eq. \eqref{eq:constraint}, and the boundary conditions
\begin{equation}\label{eq:bcccc}
        Y(0) = 0, ~~~ \theta(0) = 0, ~~~ \theta'(L) = 0,
\end{equation}
respectively prescribing the vanishing of the vertical displacement and of the rotation at the moving clamp, and the vanishing of the moment at the pin.

It should be noted that a  reaction force $R$ is realized at the pin on the profile. Due to the frictionless assumption, this force is orthogonal to the profile, and therefore inclined by the angle $\alpha$ with respect to the $\widehat X$--axis given by the following condition
\begin{equation}
    \alpha = \arctan\left[f'\left(\dfrac{d_Y}{L}\right)\right].
\end{equation}

The trivial straight configuration, 
\begin{equation}\label{eq:trivialstate}
    \theta(S)=0,\qquad\varepsilon(S)=\dfrac{P}{K},
\end{equation}
satisfies the equilibrium equations \eqref{eq:governing} if and only if $f'(0)=0$. Henceforth, this profile property is assumed and under this circumstance bifurcation may be displayed depending on the dimensionless curvature of the profile at null $\widehat{Y}$ coordinate, $f''(0)$, and rod's properties ($B$, $K$, $L$) through the stiffness ratio $q$. As shown in \cite{bigoni2012effects}, the profile shape can also be designed  to present a discontinuous curvature at null $\widehat{Y}$ coordinate, $f''(0^+)\neq f''(0^-)$ (as in Fig. \ref{fig:schematic}), to better tune the mechanical response under both signs of $\Delta$ (namely, both directions).

\section{Bifurcation and stability of the trivial configuration}  \label{sec:math_critical}
\subsection{Bifurcation loads} \label{sec:critical}
Bifurcation conditions from the trivial straight configuration, Eq. \eqref{eq:trivialstate}, are investigated under small rotation $\theta(S)$ assumption, providing the linearized version of the equilibrium equations \eqref{eq:governing}
\begin{equation} \label{eq:govern_linear}
    \begin{aligned}
        & B \theta''(S) - P\,\left[1+\varepsilon(S)\right]\,\left[\frac{d_Y}{L} f''\left(0\right)+ \theta(S) \right] = 0, \\
        & K \varepsilon(S)- P = 0,
    \end{aligned}
\end{equation}
where the following approximation is considered
\begin{equation}
    \begin{aligned}
       f'\left(\frac{d_Y}{L}\right) \approx
       \frac{d_Y}{L}f''\left(0\right).
    \end{aligned}
\end{equation}
Eq. \eqref{eq:govern_linear}$_2$ is automatically satisfied for the uniform axial strain field \eqref{eq:trivialstate}$_2$ and
 Eq. \eqref{eq:zozza}$_2$ can be linearized as
\begin{equation}
    d_Y(\varepsilon, \theta) \approx 
    \left(1+\frac{ P}{K}\right) 
    \int_0^{L} \theta(\sigma) \diff \sigma.
\end{equation}

By considering this latter approximation and the uniform axial strain field \eqref{eq:trivialstate}$_2$ and by
using the dimensionless load $p$ and the (axial vs bending) stiffness ratio $q$, defined  
by {Eq. (\ref{pera0})}, 
the equilibrium equation \eqref{eq:govern_linear}$_1$
reduces to the following homogeneous and linear integro-differential equation
\begin{equation} 
\label{eq:linearized_for_critical}
    \begin{aligned}
        &  \theta''(S) 
        -
       p\, q \left(1+p\right) \frac{\pi^2}{L^2}\left( \theta(S)+\dfrac{ f''\left(0\right)}{L}\left(1+p\right) 
    \int_0^{L} \theta(\sigma) \diff \sigma\right)
         =  
        0.
    \end{aligned}
\end{equation}
The basis for non-trivial solutions of $\theta(S)$ is found to be\footnote{The range $p<-1$ is excluded because it is representative of unphysical bifurcation, involving self-compenetrated rod configurations.}

\begin{equation} \label{eq:solution_linear}
    \theta(S)=\left\{\begin{array}{llll}
          
        a \cos\!\left(\sqrt{-(1+p)\,p\,q}
         \dfrac{\pi S}{L}\right)+
         b \sin\!\left(\sqrt{-(1+p)\,p\,q}  \dfrac{\pi S}{L}\right)+c,\qquad &p\in[-1,0],\\[4mm] 
         a \cosh\!\left(\sqrt{(p+1)\,p\,q} \dfrac{\pi S}{L}\right)+
         b \sinh\!\left(\sqrt{(p+1) \,p\,q} \dfrac{\pi S}{L}\right)+c, \qquad &p\in[0,\infty],
    \end{array}\right.
\end{equation}
where $a$, $b$, and $c$ are integration constants, where two of these can be computed as functions of the third one (which remains  arbitrary, but small, in the present analysis) by imposing the boundary conditions.

\begin{figure}
    \centering
    \includegraphics[width=80mm]{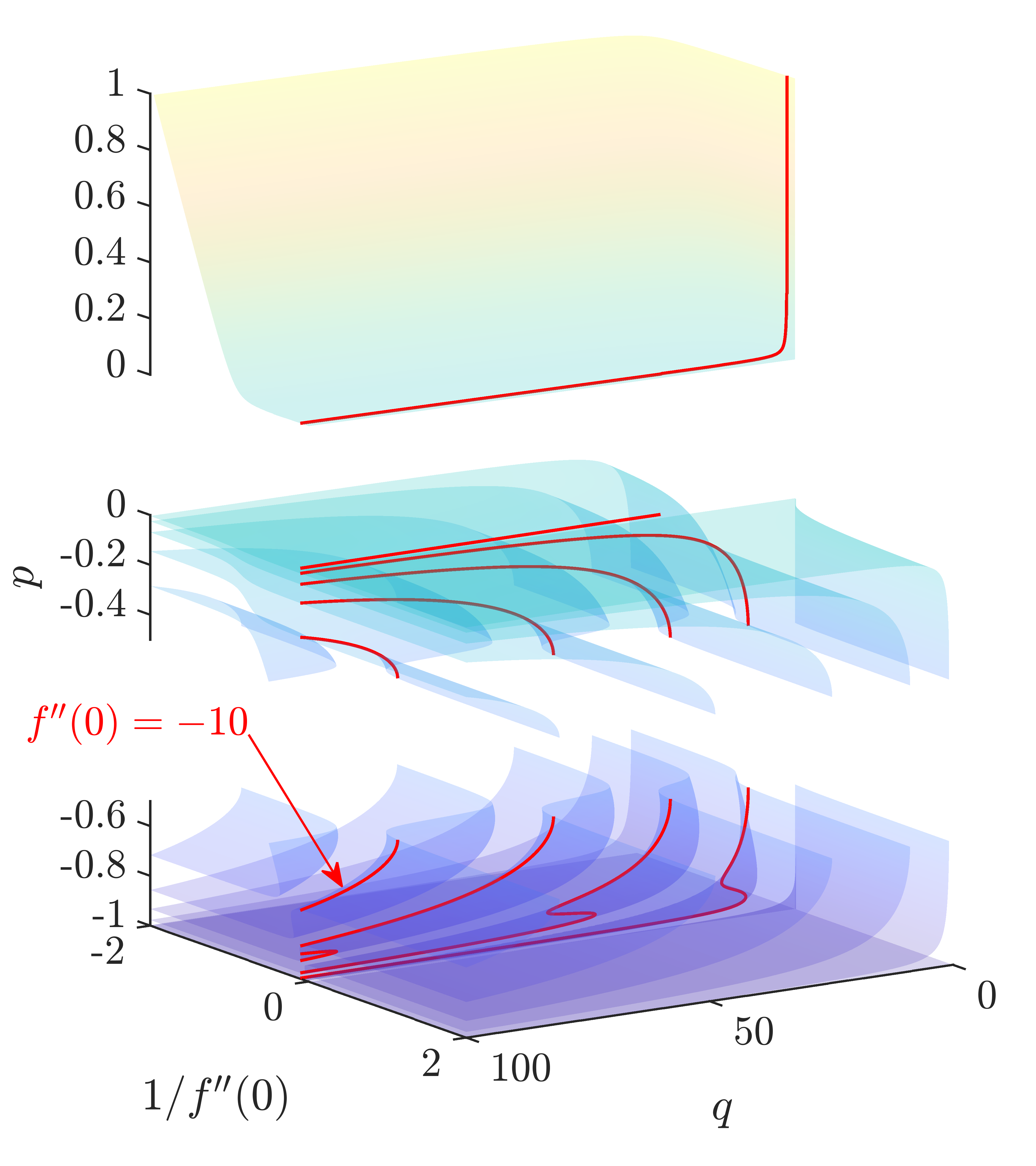}
    \caption{\label{fig:critical_surf} \footnotesize Isosurfaces of Eq. \eqref{eq:critical} showing the combination of stiffness ratio $q$, 
    dimensionless radius of profile curvature at the origin $1/f''(0)$ and load $p$ parameters for bifurcation. Note that all red curves belong to the same vertical plane defined by $1/f''(0)=-0.1$ and correspond to Fig. \ref{fig:critical_f2const} (bottom, left). The representation of the lower surfaces has been slit to improve understanding.}
\end{figure}

\begin{figure}
    \centering
    \includegraphics[width=150mm]{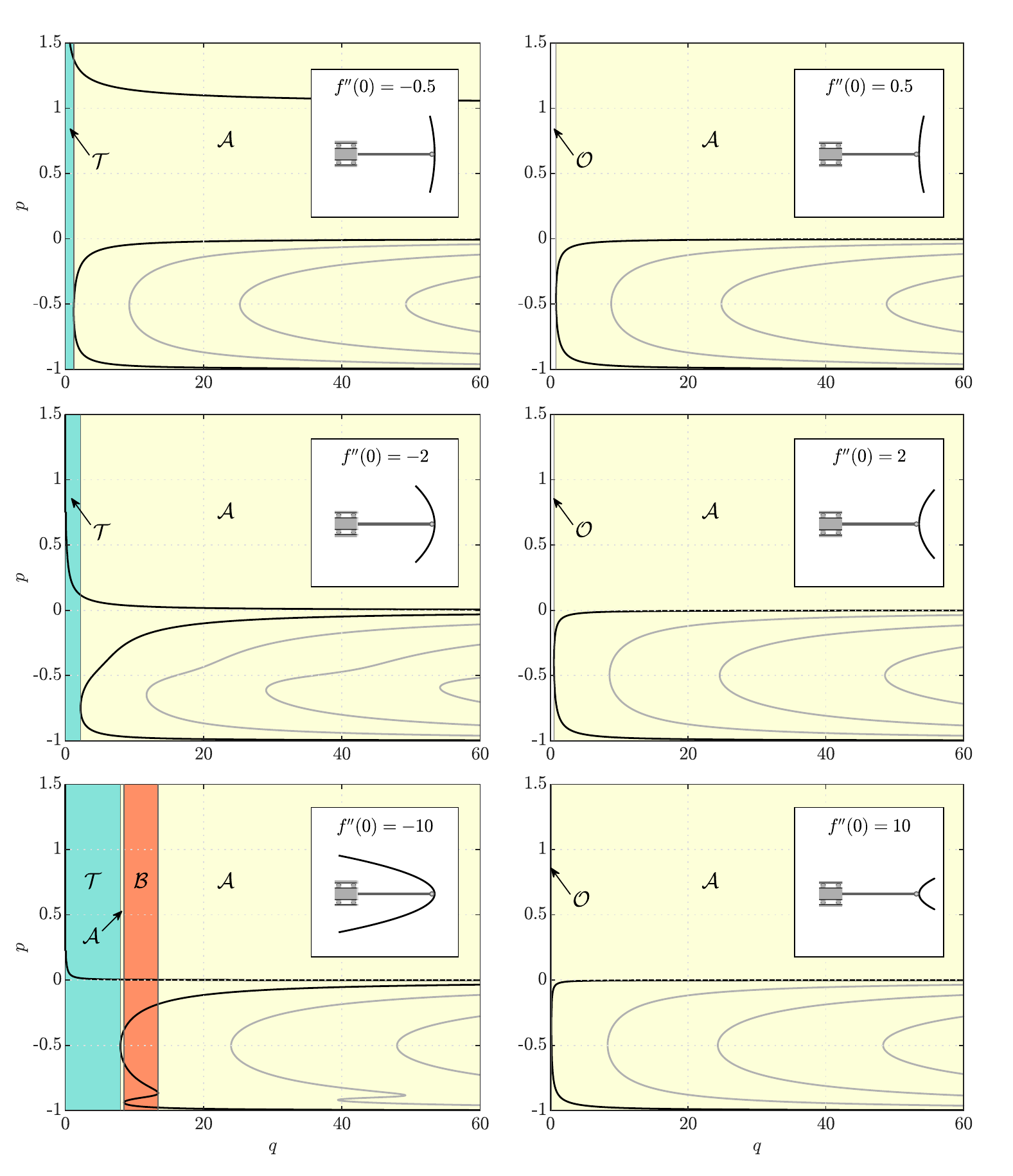}
    \caption{\label{fig:critical_f2const} 
    \footnotesize Tensile/compressive 
    bifurcation load $p$ as a function of $q$, for constant $f''(0) = \{-10,\,-2,\,-0.5,\,0.5,\,2,\,10\}$. 
    The regions $\mathcal{O}$, $\mathcal{T}$, $\mathcal{A}$, and $\mathcal{B}$ are also shown. Tensile bifurcation does not occur 
    for $f''(0)>0$, but always occurs 
    for $f''(0)<0$. In compression, multiple 
    bifurcation 
    modes can be observed and their number depends of the values of $q$ and $f''(0)$; 
    considering here the first mode, there are regions where there are none ($\mathcal{O}$ and $\mathcal{T}$), two ($\mathcal{A}$), and  four (region $\mathcal{B}$)  bifurcation compressive loads. {The grey curves represent higher modes.}
    }
\end{figure}

Therefore, by substituting solutions \eqref{eq:solution_linear} into equilibrium equation \eqref{eq:linearized_for_critical} and imposing the boundary conditions \eqref{eq:bcccc}$_2$ and \eqref{eq:bcccc}$_3$, a bifurcation with non-trivial equilibrium configuration exists when\footnote{In the inextensible limit   ($K\to\infty$, $q\to\infty$, $p\to 0$, but $p\, q= P L^2/ (\pi^2 B)$ remains finite), the bifurcation loads correspond to the solution of the following equations
\begin{equation}
    \begin{array}{ll}
         f''(0)\, \sqrt{-p\,q } + p\,q\,\pi  \, [1+f''(0)] \, \cot\left(\pi \sqrt{-p\,q  }\right) = 0, & p<0, \\[0.5em]
         f''(0)\, \sqrt{p\,q } - p\,q\,\pi\, [1+f''(0)] \, \coth\left(\pi \sqrt{p\,q  }\right) = 0, & p\geq 0.
    \end{array}
\end{equation}

The latter  equations are equivalent to that provided in \cite{bigoni2012effects} for the bifurcation loads (their Eq. (17) for $k=0$). Here, the cases of positive and negative $p$, corresponding to tension and compression respectively, are separated in order to highlight the differences between the two. The separation also facilitates calculations because developments with imaginary values can be avoided.}
\begin{equation} \label{eq:critical}
    \begin{array}{ll}
        f''(0) \sqrt{-(1+p)\, p\, q} + p\, q \,\pi\, [1+(1+p)\,f''(0)] \cot\!\left(\pi \sqrt{-(1+p)\, p\, q}\right) = 0, & p\in[-1,0], \\[0.5em]
        f''(0) \sqrt{(1+p)\, p\, q} - p\, q\, \pi\, [1+(1+p)\,f''(0)] \coth\!\left(\pi \sqrt{(1+p)\, p \,q}\right) = 0, & p\in[0,\infty),
    \end{array}
\end{equation}
defining the bifurcation load $p$ as a function of the stiffness ratio $q$ and dimensionless profile curvature at the origin $f''(0)$. 
It should be noted that the shape of the profile affects the bifurcation loads only because of its dimensionless curvature at the origin, $f''(0)$.
These bifurcation conditions are displayed in Fig. \ref{fig:critical_surf} as surfaces in the three-dimensional space defined by the parameters $p$, $q$ and $f''(0)$. Moreover, sections of these surfaces at constant values of $f''(0) = \{-10,-2,-0.5,0.5,2,10\}$  are reported 
in Fig. \ref{fig:critical_f2const}. 
%%%%%%%%%
%%%%%%%%%%%%%%%%%%%%%%%%%%%%%%%%%%%%
%%%%%%%%%%%%%%%%%%%%%%%%%%%%%%%%%%%%

%%%%%%%%%%%%%%%%%%%%%%%%%%%
From all of these representations, the following conclusions can be drawn. 
\begin{itemize}

\item Differently from a standard Sturm-Liouville problem, governing bifurcation for an  inextensible rod ($q\rightarrow\infty$),  the number of bifurcation loads and modes becomes finite due to rod's extensibility. For instance, in Fig. \ref{fig:critical_f2const} (upper part on the left) for stiffness ratio $q<1.231$ there is only one mode in tension, while for $q\in(1.231,\,9.264)$ there is  only one mode in tension and only one in compression. Moreover, in Fig. \ref{fig:critical_f2const} (upper part on the right) there is only one mode in compression for stiffness ratio $q\in(0.822,\,8.836)$.

\item Another effect related to the axial compressibility of the rod is that a  bifurcation {\it mode} in compression corresponds to more than one bifurcation load. Indeed, except in the limit cases of coalescent loads (discussed below), the bifurcation loads in compression always occur {\it in pairs}, so that one corresponds to the \emph{destabilization} (loss of stability) of the straight configuration, while the other to its \emph{restabilization}. For instance, in Fig. \ref{fig:critical_f2const} (upper part on the left) for $q>1.231$ there is one bifurcation load for the first mode corresponding to the transition from stability to instability (at increasing 
compressive load) of the straight configuration. In addition, one restabilization load is found for the first mode, corresponding to the transition from instability to stability. Moreover, it can be observed in the same figure that for higher stiffness ratio $q$, where a plurality of modes emerge, the compressive bifurcation loads always occur in pairs.

    \item In tension: none or only one bifurcation load may exist, depending on the sign of the profile curvature at the origin  $f''(0)$. 
    The fact that only one critical load is found is related to the presence of the hyperbolic cotangent  function in the bifurcation equation \eqref{eq:critical}$_2$.
    In particular:
    \begin{itemize}
        \item for $f''(0)>0$, tensile bifurcation does not occur, as in  the inextensible case;
        \item for $f''(0)<0$, one tensile bifurcation occurs at $p_{cr}^{(+)}>0$ (where the superscript $(+)$ reminds the reference to tensile load). This is in contrast with the inextensible case, where tensile bifurcation was  existing in a more limited set, $f''(0)<-1$~\cite{bigoni2012effects}.
    \end{itemize}
    \item In compression: either bifurcation is excluded, or it may occur with 2 or 4  bifurcation loads associated to the same mode, depending on the values of $q$ and $f''(0)$.
    More specifically, restricting attention only to the first compression mode, 2 main cases  can be distinguished, differing in the number of bifurcations at varying $q$ and $f''(0)$:
    \begin{itemize}
    \item%[$\mathcal{A}$ --] 
    2 bifurcation loads occur, associated to the first mode. Each of these corresponds to a destabilization  load $\left(p_{de}^{(-)}\right)$ and a restabilization load $\left(p_{re}^{(-)}\right)$  (where the superscript $(-)$ reminds the reference to compressive load), 
    \beq
    -1<p_{re}^{(-)}<p_{de}^{(-)}<0.
    \eeq
    An example of this behaviour is visible in Fig. \ref{fig:critical_f2const} in the upper part on the left for $q>1.231$;
    \item%[$\mathcal{B}$ --] 
    4 bifurcation loads occur, associated to the first mode, corresponding to two 
    pairs of destabilization/restabilization loads, 
    $\left(p_{de,1}^{(-)}, p_{re,1}^{(-)}\right)$  and $\left(p_{de,2}^{(-)}, p_{re,2}^{(-)}\right)$, occurring in the following order\footnote{
  In the case when double restabilization may occur for a value of $f''(0)$, the nomenclature of the bifurcation loads $p_{de}^{(-)}$ and $p_{re}^{(-)}$ is enhanced by introducing a subscript 1 or 2 to distinguish between the first and second destabilization/restabilization loads. In order to simplify the presentation, this notation is preserved through a continuation principle for the roots passing from a set of $q$ values where double restabilization is displayed to the set of single restabilization, as illustrated in Fig. \ref{fig:pinned}.
  }
    \beq
    -1<p_{re,2}^{(-)}<p_{de,2}^{(-)}<p_{re,1}^{(-)}<p_{de,1}^{(-)}<0.
    \eeq
    An example of this behaviour is visible in Fig. \ref{fig:critical_f2const} in the lower part on the left $q\in(8.488,13.451)$;
    \end{itemize}
   
   \item In compression: multiple bifurcation loads associated to the same mode may occur for modes higher than the first. An example is visible in Fig. \ref{fig:critical_f2const} in the lower part on the left.
\end{itemize}

\begin{figure}
    \centering
    \includegraphics[width=150mm]{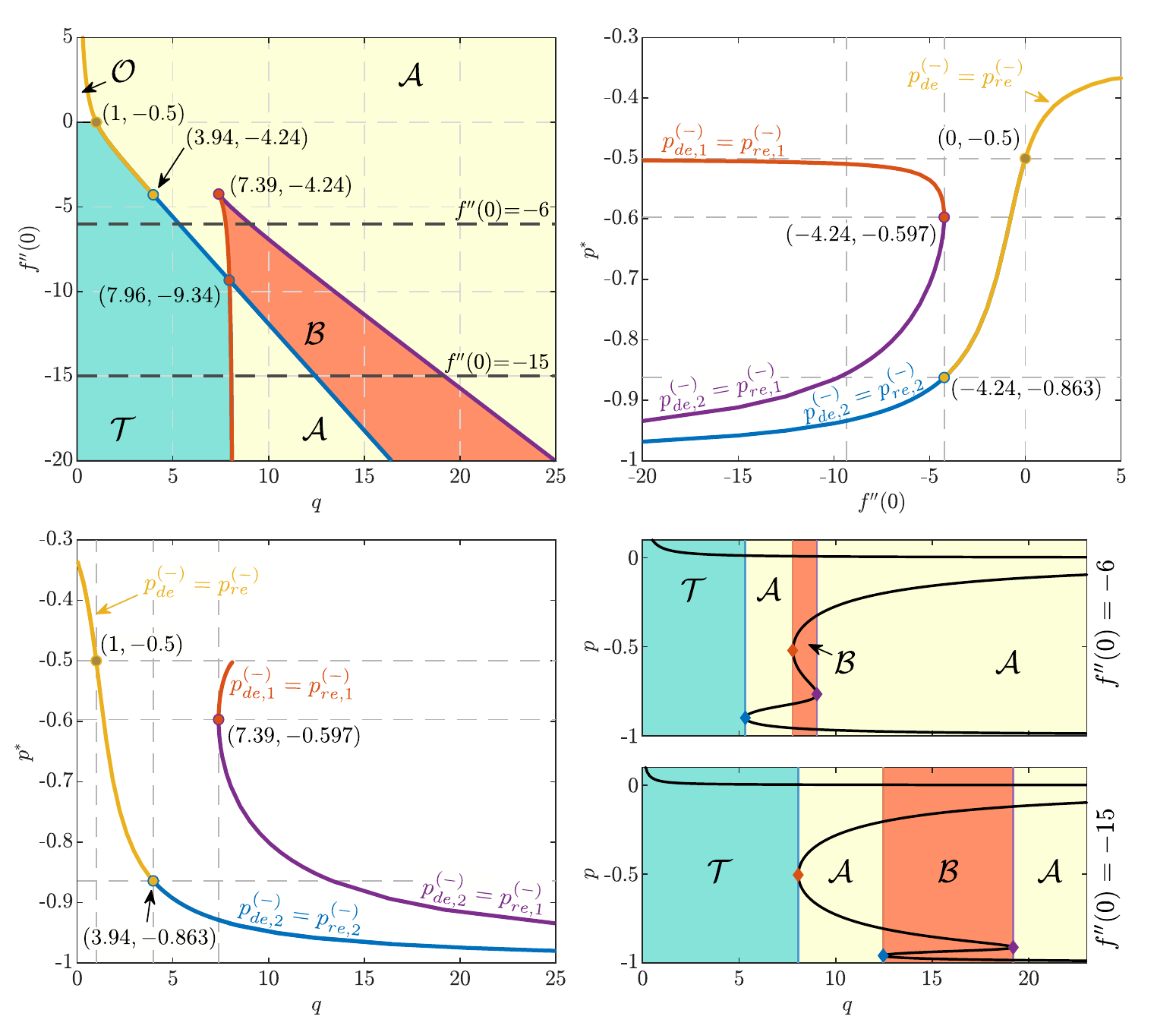}
    \caption{\label{fig:critical_regions} \footnotesize Upper part, left: Sets $\mathcal{O}$, $\mathcal{T}$, $\mathcal{A}$, and $\mathcal{B}$ defined by the number of critical loads in compression (restricted to first mode) and in tension, as described in  Table \ref{tab:otab}.  
    {Upper part, right and lower part, left: }
    Coalescent bifurcation load $p^*$ as a function of $q$ (left, lower part) and of $f''(0)$  (right, upper part), defining the different boundaries between two 
    of the sets
    $\mathcal{O}$, $\mathcal{T}$, $\mathcal{A}$, and $\mathcal{B}$. Lower part, right: Bifurcation load $p$ versus $q$, evaluated for two values of $f''(0)$; here $p^*$ is marked with a diamond of the same colour as the corresponding transition curve.
    }
\end{figure}

By considering all of the above described cases, four different  subsets $\mathcal{O}$, $\mathcal{T}$, $\mathcal{A}$, and $\mathcal{B}$
    are distinguished in the $q-f''(0)$ plane (Fig. \ref{fig:critical_regions}, upper part, on the  left), differing in the number of the main bifurcation conditions as reported in Tab. 1.
    \begin{table}[H]
    \caption{\label{tab:otab} Numbers of main bifurcations for the  subsets $\mathcal{O}$, $\mathcal{T}$, $\mathcal{A}$, and $\mathcal{B}$.}
    \centering
    \begin{tabular}{c|c|c}
         & Tensile bifurcation & Compressive bifurcation \\\hline
        $\mathcal{O}$ & 0  &  0 \\\hline
        $\mathcal{T}$ & 1  &  0 \\\hline
        $\mathcal{A}$ & 1  &  2 bif. loads associated with $1^{\mbox{\tiny {st}}}$ mode \\\hline
        $\mathcal{B}$ & 1  &  4 bif. loads associated with $1^{\mbox{\tiny{st}}}$ mode
    \end{tabular}
\end{table}

    In the following description of the bifurcation response, the terms 
    \lq \emph{destabilization}' and 
    \lq \emph{restabilization}' are introduced to respectively define a bifurcation load in compression representing a transition for the trivial configuration to become unstable or to return to being stable for compressive loads at increasing magnitude. 
    
    The $q-f''(0)$ pairs, defining the boundary of each subset, represent the transition in the bifurcation response. 
        More specifically:
    \begin{itemize}
           \item the transition from subset $\mathcal{A}$ to $\mathcal{O}$ or $\mathcal{T}$ 
           corresponds to the transition from the presence of a destabilization
           $p_{de}^{(-)}$
           and restabilization 
           $p_{re}^{(-)}$ load 
        (belonging to the subset $\mathcal{A}$)
           to a situation where bifurcation is excluded. This occurs at the coalescence
           \begin{equation}
           \label{palle2}
 p_{de}^{(-)}=p_{re}^{(-)};
     \end{equation}
           \item the transition from subset $\mathcal{A}$ to $\mathcal{B}$ correspond to the transition from 
           a pair of bifurcation loads $\left(p_{de}^{(-)}, p_{re}^{(-)}\right)$ to two pairs 
           $\left(p_{de,1}^{(-)}, p_{re,1}^{(-)}\right)$ and 
           $\left(p_{de,2}^{(-)}, p_{re,2}^{(-)}\right)$, 
 all associated to the first 
           bifurcation 
           mode in compression.
           This necessarily involves one of the following coalescences
\begin{equation}
      \label{palle3}
        p_{de,1}^{(-)}=p_{re,1}^{(-)},      
\qquad \mbox{or}\qquad
     p_{de,2}^{(-)}=p_{re,2}^{(-)}, \qquad \mbox{or}\qquad 
     p_{de,2}^{(-)}=p_{re,1}^{(-)}. 
\end{equation}     
\end{itemize}
Defining the range of $q\in(q_a,q_b)$ limiting  the $\mathcal B$ region, the bound values $q_a$ and $q_b$ are approximately linear in the dimensionless profile curvature when this takes large negative values, \begin{equation}
    q_a\approx 0.526-0.796 f''(0)\,,\quad q_b\approx 1.655-1.167 f''(0),
    \qquad
    \mbox{when}\,\,f''(0)< -10.
\end{equation}

    It is anticipated from Section \ref{sec:vibr} 
that,  
considering also tensile buckling
(occurring at $p_{cr}^{(+)}\geq 0$), 
the set $\mathcal{A}$ 
can be divided into 
two pairs of subsets, so that 
when the load $p$ belongs to one pair  
the trivial configuration is stable, otherwise is unstable
\beq    
\begin{array}{lll}\label{stableA}
\mbox{if}\,\,     p\in
     \left(-1,p_{re}^{(-)}\right)
      \cup 
      \left(p_{de}^{(-)},p_{cr}^{(+)}\right)
      & \Rightarrow &\mbox{trivial configuration is stable},\\[4mm]
\mbox{if}\,\,     p\in\left(p_{re}^{(-)},p_{de}^{(-)}\right) \cup \left(p_{cr}^{(+)},\infty\right) &  \Rightarrow &\mbox{trivial  configuration is unstable}, 
\end{array}
\eeq
while 
the set $\mathcal{B}$ 
can be divided into two triplets 
of subsets 
\beq  
\begin{array}{lll}\label{stableB}
\mbox{if}\,\,      p\in
     \left(-1,p_{re,2}^{(-)}\right)
      \cup 
      \left(p_{de,2}^{(-)},p_{re,1}^{(-)}\right)
      \cup 
      \left(p_{de,1}^{(-)},p_{cr}^{(+)}\right)
      &  \Rightarrow &\mbox{\emph{stable} trivial conf.,}\\[4mm]
\mbox{if}\,\,      p\in\left(p_{re,2}^{(-)},p_{de,2}^{(-)}\right) \cup\left(p_{re,1}^{(-)},p_{de,1}^{(-)}\right) \cup \left(p_{cr}^{(+)},\infty\right) & \Rightarrow & \mbox{\emph{unstable} trivial conf.}
\end{array}
\eeq

    The conditions of coalescence correspond to 
    triplets of values for the parameters $p$, $q$, and
    $f''(0)$ and can be visualized as a curve 
    in a three-dimensional parameter space. To investigate coalescence in a two-dimensional representation, Fig.  \ref{fig:critical_regions} is introduced. 
In the upper part (on the left) of this figure, all  transitions between the different sets 
$\mathcal{O}$, $\mathcal{T}$, $\mathcal{A}$, and $\mathcal{B}$
are reported 
in the $q-f''(0)$ plane as curves drawn with different colours. 
    The bifurcation load at which coalescence occurs is indicated with $p^*$ and is reported in Fig.  \ref{fig:critical_regions} at varying $q$ 
    (lower part, on the left) and at varying $f''(0)$     (upper part, on the right). Note that in the former case $f''(0)$ 
    (in the latter case $q$) 
    does not assume a fixed value, but varies. 
    
    A detail of the bifurcation $p$-$q$ curves at fixed
    $f''(0)=\{-6, -15\}$ is reported in the lower part (on the right) of Fig. \ref{fig:critical_regions}. Here, occurrences of coalescence $p=p^*$ are marked with diamonds of the same colour as the corresponding transition
curve.

% %%%%%%%%%
% \begin{figure}
%     \centering
%     \includegraphics[width=80mm]{figures/critical_iso_surf_cut.png}
%     \caption{\label{fig:critical_surf} Isosurfaces of Eq. \eqref{eq:critical} showing the combination of stiffness ratio $q$, 
%     constraint curvature $1/f''(0)$ and load $p$ parameters for bifurcation. Note that all red curves belong to the same vertical plane defined by $1/f''(0)=-0.1$ and correspond to Fig. \ref{fig:critical_f2const} (bottom, left). The representation of the lower surfaces has been slit to improve understanding.}
% \end{figure}
% %%%%%%%%%%%%%%%%%%%%%%%%%%%%%%%%%%%%

% %%%%%%%%%%%%%%%%%%%%%%%%%%%%%%%%%%%%
% \begin{figure}
%     \centering
%     \includegraphics[width=150mm]{figures/qeffect.eps}
%     \caption{\label{fig:critical_f2const} 
%     Tensile/compressive 
%     bifurcation load $p$ as a function of $q$, for constant $f''(0) = \{-10,\,-2,\,-0.5,\,0.5,\,2,\,10\}$. 
%     %
%     The regions $\mathcal{O}$, $\mathcal{T}$, $\mathcal{A}$, and $\mathcal{B}$ are also shown. Tensile bifurcation does not occur 
%     for $f''(0)>0$, but always occurs 
%     for $f''(0)<0$. In compression, multiple 
%     bifurcation 
%     modes can be observed and their number depends of the values of $q$ and $f''(0)$; 
%     considering here the first mode, there are regions where there are none ($\mathcal{O}$ and $\mathcal{T}$), two ($\mathcal{A}$), and  four (region $\mathcal{B}$)  bifurcation loads. 
%     }
% \end{figure}

% %%%%%%%%%%%%%%%%%%%%%%%%%%%

%%%%%%%%%%%%%%%%%%%%%%%
%%%%%%%%%%%%%%%%%%%%%%%%%%

\begin{figure}
    \centering
    \includegraphics[width=155mm]{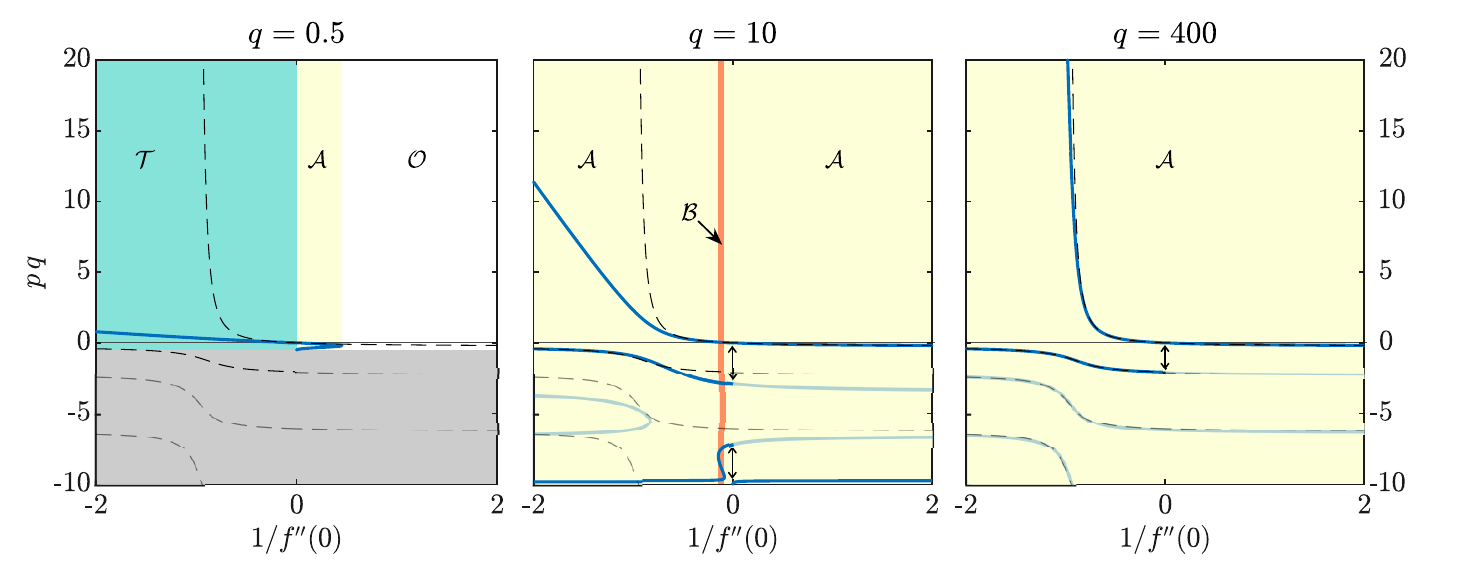}
    \caption{\label{fig:critical_qconst} 
    \footnotesize Comparison between tensile/compressive bifurcation loads $pq=P L^2/(\pi^2 B)$ for extensible (dashed lines) and inextensible (continuous lines) rod models and for $q=\{0.5,10,400\}$. 
    The region shaded in gray corresponds to $p<-1$, which has no mechanical meaning. 
    Bifurcation loads corresponding to modes higher than the first are drawn with lighter lines. 
    As the stiffness ratio $q$ increases, the bifurcation models converge. 
    At the transition point $1/f''(0)=0$ tensile instability turns to compressive and the critical load suffers a jump. 
    }
\end{figure}

A comparison is presented in Fig. \ref{fig:critical_qconst} between bifurcations calculated with two models 
of elastic rod, namely, an inextensible 
Euler-Bernoulli rod, and 
the axially deformable rod 
investigated in this article, 
both 
characterized by the same 
bending stiffness $B$ and the latter investigated for constant values of $q = \{0.5, 10, 400\}$.

The comparison is reported in terms of a dimensionless value of the axial load $P$, namely, $pq = PL^2/(\pi^2 B)$,
reported as a function of the dimensionless radius of profile curvature at the origin $1/f''(0)$. 

Note 
in the figure 
that: (i.) tensile and compressive loads are considered; (ii.) some modes higher than the first are included (in light colour); (iii.)  
the response of the inextensible (extensible) model is reported 
with a dashed (continuous) line; 
(iv.) the region $p<-1$, highlighted though a gray background, should not be considered as the related compression level has no physical meaning.

From this figure, two main conclusions can be drawn:
\begin{itemize}
    \item as the stiffness ratio $q$  increases (from left to right in the figure), the bifurcation behaviour of the extensible rod  converges to that pertaining to the inextensible one;

    \item 
in the figure on the left both models show a transition from a tensile to a compressive bifurcation load (the latter terminates when $p=-1$) and other bifurcation modes are not available. Even in both figures pertaining to $q= 10$ and $400$, and for both rod models, 
a tensile bifurcation load becomes compressive 
at the point $1/f''(0)=0$. However, defined as the smaller (in absolute value) load, the critical load 
evidences a jump at this point, because 
the continuation of the tensile critical load prevails on a higher-order mode, which is critical for $1/f''(0)<0$. 

\end{itemize}
For completeness, the regions $\mathcal{O}$, $\mathcal{T}$, $\mathcal{A}$, and $\mathcal{B}$ are also shown in Fig. \ref{fig:critical_qconst} and these regions suggest the following observations.

\begin{itemize}
   \item  The compressive critical load $p_{de}^{(-)}$ for positive profile curvature $f''(0)=f''_a$  is bounded  as follows (Fig. \ref{fig:critical_regions})
\beq
p_{de}^{(-)}\left(f''_a,q\right)\in\left[-\dfrac{1}{2},0\right],\qquad 
\forall\,\, f''_a>0 .
\eeq

 \item With reference to two values  $f''_a$ and $f''_b$ of profile curvature at the origin $f''(0)$, the following inequalities hold for the critical loads in tension (Fig. \ref{fig:critical_qconst})
\beq
f''_a<0 \,\,\, \mbox{ and } \,\,\,
f''_a<f''_b
\qquad \Leftrightarrow\qquad
p_{cr}^{(+)}\left(f''_a,q\right) \leq p_{cr}^{(+)}\left(f''_b,q\right),\qquad\forall\,\,q >0,
\eeq
and in compression
\beq
\label{bifloadprop}
f''_a<f''_b\qquad \Rightarrow\qquad 
p_{de,1}^{(-)}\left(f''_a,q\right) < p_{de,1}^{(-)}\left(f''_b,q\right),
\eeq
\end{itemize}
where the definition of 
 critical loads is extended to include
  cases in which 
   bifurcation does not occur, by  assuming   
 $p_{cr}^{(+)}\left(f'',q\right)=+\infty$ for $\{f'',q\}\in\mathcal{O}$ and
$p_{de,1}^{(-)}\left(f'',q\right)=-\infty$ for $\{f'',q\}\in\mathcal{O}\cup\mathcal{T}$.
%%%%%%%%%%%%%%%%%%%%%%%%%%%%%%%%%%%%
%%%%%%%%%%%%%%%%%%%%%%%%%%%%%%%%%%%%

Asymptotic expressions for the bifurcation loads in the limit of vanishing profile curvatures, $f''(0)\to0^\pm$, are reported in Appendix \ref{appendixasymptotic}.

\paragraph{Profile with curvature discontinuity at the origin.}\label{discprof}
When the profile possesses a curvature discontinuity at the origin, $f''\left(0^+\right)\neq f''\left(0^-\right)$, 
each part of the profile defines its \lq own' critical loads in tension $p_{cr}^{(+)}\!\left(f''\!\left(0^\pm\right)\!,\,q\right)$ and in compression $p_{de}^{(-)}\!\left(f''\!\left(0^\pm\right)\!,\,q\right)$. The tensile (compressive) critical load for the structure 
corresponds to the smallest (the highest) load, 
\beq\label{ppcrcr}
\begin{array}{llll}
p_{cr}^{(+)}=\min\left[p_{cr}^{(+)}\!\left(f''\!\left(0^+\right)\!,\,q\right)\!,\,p_{cr}^{(+)}\!\left(f''\!\left(0^-\right)\!,\,q\right)\right]>0,\\[4mm]
p_{cr}^{(-)}=\max\left[p_{de,1}^{(-)}\!\left(f''\!\left(0^+\right)\!,\,q\right)\!,\,p_{de,1}^{(-)}\!\left(f''\!\left(0^-\right)\!,\,q\right)\right]<0.
\end{array}
\eeq
With the above premise, the design proposed in this article for the profile  is based 
on the following two requirements.
\begin{itemize}
    \item 
The first design requirement is a prescription  that the shape of the profile that leads to a  postcritical behaviour in tension and in compression is of the type illustrated in 
Fig. \ref{fig:schematic}, so that in tension 
$
p_{cr}^{(+)}=p_{cr}^{(+)}\!\left(f''\!\left(0^-\right)\!,\,q\right)
$
and in compression
$
p_{cr}^{(-)}=p_{de,1}^{(-)}\!\left(f''\!\left(0^+\right)\!,\,q\right)
$. From a kinematical point of view, this requirement is equivalent to the 
prescription that 
bifurcation under compression occurs with $d_Y>0$ and under tension with $d_Y<0$. When these conditions prevail, the following inequality holds
\beq\label{deltap}
\Delta(p)\,d_Y(p)\leq 0,
\eeq
and  the critical loads \eqref{ppcrcr} become only functions of $f''(0)$ and $q$
\beq
\label{pcrchosen}
   p_{cr}^{(+)}=p_{cr}^{(+)}\!\left(f''\!\left(0^-\right)\!,\,q\right),\qquad
   p_{cr}^{(-)}=p_{cr}^{(-)}\!\left(f''\!\left(0^+\right)\!,\,q\right).
\eeq

\item The second design requirement is that the values of the two critical loads  $p_{cr}^{(+)}$ and $p_{cr}^{(-)}$ must be assigned as desired.

\end{itemize}

It is clear that the inequality (\ref{deltap}) 
imposes a 
restriction on the possibility to arbitrarily assign the two critical loads. This restriction can be obtained as follows.

After the loads are assigned, it can be proven that a specific value of $q$ exists, say, $\bar{q}$, for which the critical loads are obtained with the same profile curvature at the origin, $f''(0^-)=f''(0^+)$, say, $\bar{f}''$. 

For $q \geq \bar{q}$ the inequality (\ref{deltap}) fails, because 
\begin{equation}
    q > \bar{q} \,\,\, \Longrightarrow
    \,\,\,  f''(0^+) > f''(0^-) \,\,\,\, \mbox{ and } \,\,\,\,
    p_{de,1}^{(-)}\left(f''(0^-)\right) > 
    p_{de,1}^{(-)}\left(f''(0^+)\right)
\end{equation}
meaning that the rod buckles in compression ($p<0$) with $d_Y<0$, oppositely to what is assumed through the inequality \eqref{deltap}. More specifically, for given values of $p_{cr}^{(+)}$ and $p_{cr}^{(-)}$, the pairs $\left\{\bar f'',\, \bar q\right\}$ define the range of values of $f''(0^+)$, $f''(0^-)$ and $q$ for which Eq. (\ref{deltap}) holds as 
\beq\label{ranges}
f''\!\left(0^-\right)\leq\bar f''\leq 0 \leq f''\!\left(0^+\right),
\qquad
q\leq\bar q. 
\eeq

\begin{figure}[t]
    \centering
    \includegraphics[width=155mm]{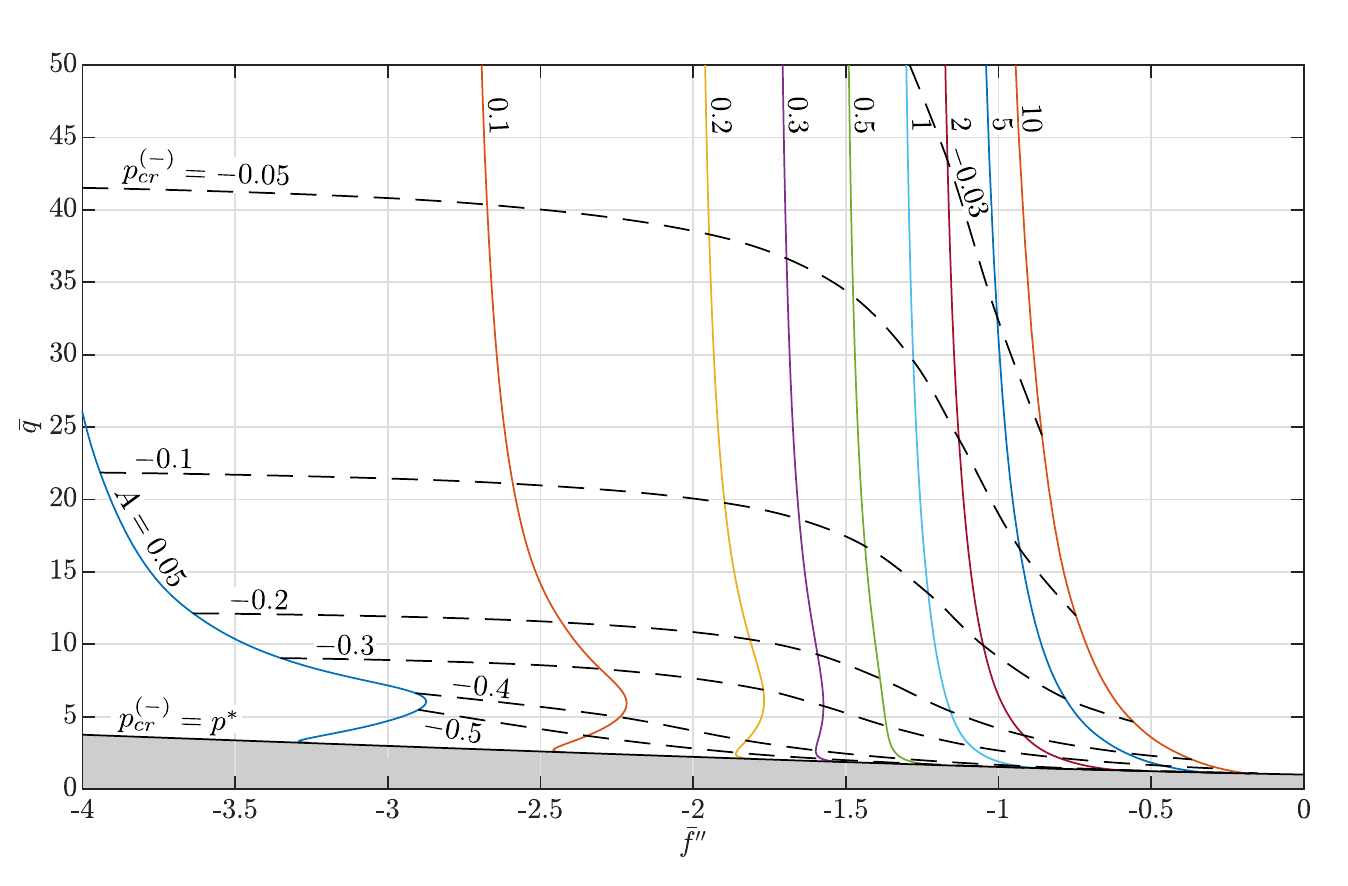}
    \caption{\label{fig:qbar} \footnotesize Parametric 
    representation of profile curvature 
    at the origin $\bar f''$ 
    and stiffness ratio $\bar q$ defining the limit parameters for which the two critical loads can be tuned separately. 
    The intersection between (solid and coloured) curves at different $A=-p_{cr}^{(+)}/p_{cr}^{(-)}$ 
    and (dashed) curves pertaining to 
    $p_{cr}^{(-)}$ defines the corresponding pair 
    ($\bar{f}''$, $\bar{q}$). 
    In the gray zone compressive bifurcation does not occur  and its boundary corresponds to coalescent bifurcation loads $p_{cr}^{(-)}=p^*$. }
\end{figure}

Introducing the ratio $A=-p_{cr}^{(+)}/p_{cr}^{(-)}$, the
 parametric curves $\left\{\bar f''\!\left( p_{cr}^{(-)}\!,\,A\right), \bar{q}\left( p_{cr}^{(-)}\!,\,A\right)\right\}$ are reported in Fig. \ref{fig:qbar}. The points at the intersection between a (dashed) curve pertaining to a value of 
  $p_{cr}^{(-)}$ and a (coloured, solid) curve 
  representative of a value of $A$ lead the values of the pair $\bar{f}''$ and $\bar{q}$ reported on the axes.  Note that inside the gray zone in the figure bifurcation is excluded. The boundary of this zone corresponds to the occurrence of  coalescent bifurcation loads $p_{cr}^{(-)}=p^*$.

\subsection{Stability of the straight configuration from small amplitude vibration analysis} \label{sec:vibr}

The stability of the straight configuration can be assessed by analysing the nature of the frequency of the  time-harmonic small amplitude vibrations around the straight configuration. In particular, the presence (or absence) of non-real eigenfrequencies defines the instability (or stability) of the straight configuration, and the limit case of vanishing eigenfrequency would confirm the  critical loads obtained from the quasi-static analysis performed in Sect. \ref{sec:critical}.
With the purpose of studying the dynamic response of the system, a linear mass density $\gamma$ is considered for the rod, so that  
 the Lagrangian $\mathcal{L}$ of the system can be written as
\begin{equation}
    \begin{multlined}
        \mathcal{L} = \int_0^L \frac{\gamma}{2} \left(\dot {\widehat X}^2 + \dot {\widehat Y}^2\right) \diff S 
        - \int_0^L \frac{B}{2} \left(\theta'\right)^2 \diff S 
        - \int_0^L \frac{K}{2} \varepsilon^2 \diff S
        - \int_0^L N_X \left\{ \widehat X' - (1+\varepsilon) \cos\theta \right\} \diff S \\
        - \int_0^L N_Y \left\{ \widehat Y' - (1+\varepsilon) \sin\theta \right\} \diff S
        - P\, \widehat X(0) 
        + R_X \left\{\widehat X(L)-L\left[f\!\left(\frac{\widehat Y(L)}{L}\right)\right]\right\},
    \end{multlined}
\end{equation}
where $N_X$, $N_Y$, and $R_X$ are Lagrangian multipliers (the first two representing the $X$ and $Y$ components of the internal force,  the last representing the $X$ component of the  reaction $R$ at the pin on the curved profile) and the dot over the functions stands for differentiation with respect to the time variable $t$.
The equations of motion can be derived from the principle of least action as
\begin{equation} \label{eq:vibr_eom}
    \begin{aligned}
        & N_X'=\gamma\,\ddot {\widehat X} , \\
        & N_Y'=\gamma\,\ddot {\widehat Y} , \\
        & B\, \theta'' - (1+\varepsilon) [N_X\, \sin\theta - N_Y\, \cos\theta] = 0, \\
        & K\, \varepsilon - N_X\,\cos\theta - N_Y\,\sin\theta = 0,
    \end{aligned}
\end{equation}
along with the boundary conditions
\begin{equation}\label{eq:vibr_bc}
 N_X(0) = P , \qquad
         N_X(L) = R_X,  \qquad
         N_Y(L) = R_X\,f'\left(\frac{\widehat Y(L)}{L}\right).
 \end{equation}
Considering small time-harmonic vibrations { \cite{bosi2016}
} around the trivial equilibrium configuration \eqref{eq:trivialstate}, the  relevant fields are assumed through the method of separation of variables as
\begin{equation}
    \begin{aligned}
        & \widehat X(S,t) = S + \frac{P}{K} (S-L) + \bar X(S)\,e^{i\,\omega_x t}, \qquad & \widehat Y(S,t) = \bar Y(S)\,e^{i\,\omega_y t}, \\
        & \varepsilon(S,t) = \frac{P}{K} + \bar\varepsilon(S)\, e^{i\,\omega_x t}, \qquad
        & \theta(S,t) = \bar\theta(S)\,e^{i\,\omega_y t}, \\
        & N_X(S,t) = P + \bar N_X(S)\,e^{i\,\omega_x t},\qquad & N_Y(S,t) = \bar N_Y(S)\,e^{i\,\omega_y t}, \\
        & R_X(t) = P + \bar R_X\,e^{i\,\omega_x t},
    \end{aligned}
\end{equation}
where $\omega_x$ and $\omega_y$ are the circular frequencies for axial and flexural vibrations, respectively,  and functions with an overbar denote functions of the { spatial} variable only. The amplitudes of the overbar functions are considered to be driven by a small positive parameter $\eta$, so that
\beq
    |\bar X| \approx |\bar Y| \approx |\bar\varepsilon| \approx |\bar\theta| \approx |\bar N_X| \approx |\bar N_Y| \approx |\bar R_X| \approx \eta,
\qquad
    \eta \gg \eta^2.
\eeq

\begin{figure}
    \centering
    \includegraphics[width=150mm]{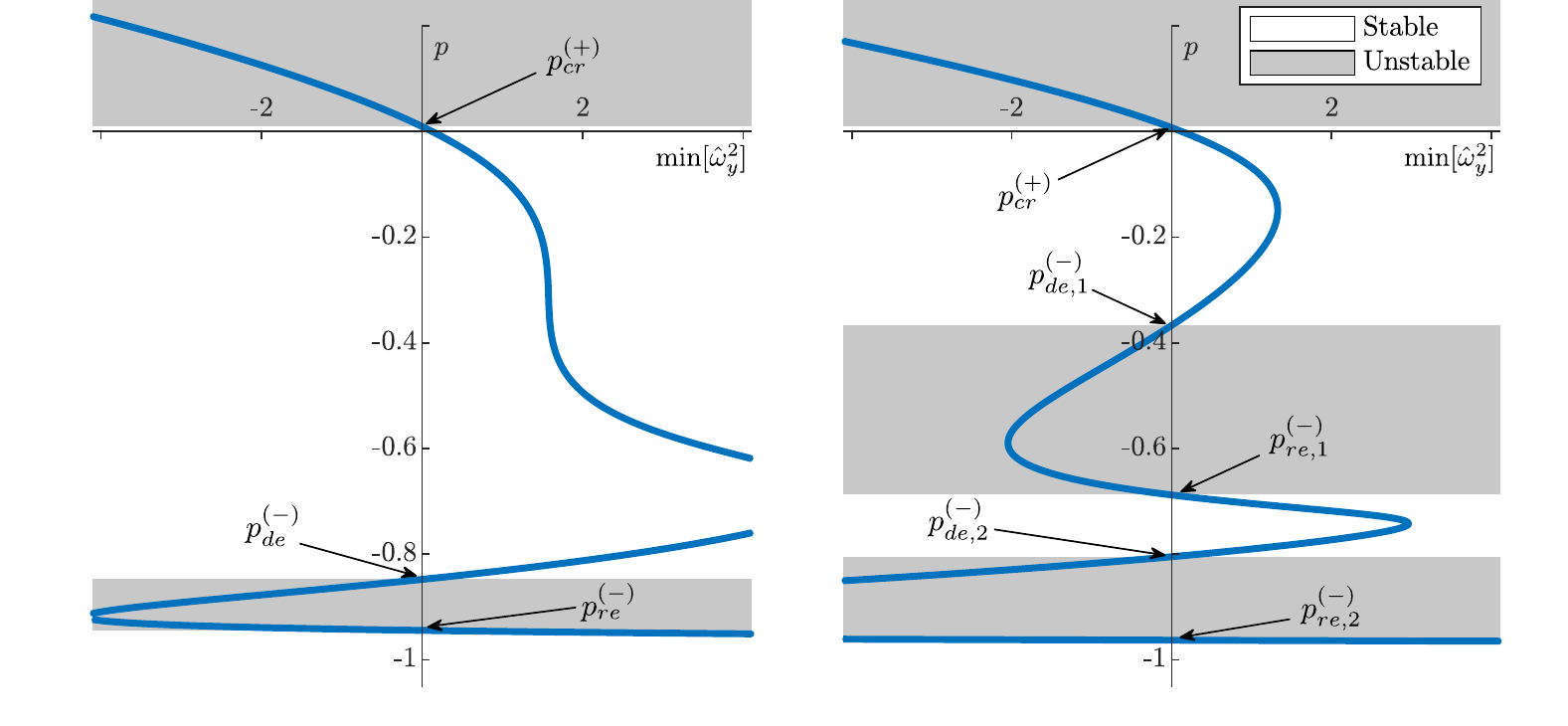}
    \caption{\label{fig:eigenfreq} 
    \footnotesize Smallest eigenvalue  $\hat{\omega}_y^2=\gamma\,\omega_y^2 L^4/(\pi^4 B)$  for $f''(0)=-6$ when the dimensionless applied load $p$ is varied. Two stiffness ratios are considered to analyse the stability in subset $\mathcal{A}$ (left, $q=6.5$) and subset $\mathcal{B}$ (right, $q=8.5$).  Stability (or instability) of the straight configuration corresponds to positiveness (or negativeness) of the smallest eigenvalue $\hat{\omega}_y^2$. Restabilization (left) or double restabilization (right) is found. }
\end{figure}

A substitution of the above expressions into Eqs. \eqref{eq:vibr_eom} and \eqref{eq:vibr_bc}, by neglecting higher-order terms in  $\eta$, yields the following decoupled differential equations in the axial and transverse amplitude oscillations
\begin{equation} \label{eq:decoupled_osc}
    \begin{aligned}
        & \bar X''(S) + \hat{\omega}_x^2\frac{\pi^2}{L^2}\,\bar X(S) = 0 , \\
        & {\bar Y}^{\mbox{\tiny (iv)}}(S) - (1+p)\,p\,q\,\frac{\pi^2}{L^2}\, \bar Y''(S) - (1+p)^2\, \hat{\omega}_y^2\frac{\pi^4}{L^4}\, \bar Y(S) = 0 ,
    \end{aligned}
\end{equation}
 where
\begin{equation}
    \hat{\omega}_x^2 = \frac{\gamma\, \omega_x^2 L^2}{\pi^2 K},\quad \hat{\omega}_y^2 = \frac{\gamma\,\omega_y^2 L^4}{\pi^4 B} ,
\end{equation}
to be complemented with the boundary conditions\footnote{The  expansion of Eq. \eqref{eq:constraintprofile} truncated to linear terms in  $\widehat Y(L)$ is
\beq
    \widehat X(L)=L f(0) + f'(0)\widehat Y(L), 
    \eeq
which, under the considered assumption  $f'(0)=0$, reduces to
\beq
    \widehat X(L) = L f(0),
\eeq
implying the boundary condition (\ref{eq:boundary_cond})$_2$.
}
\begin{equation} \label{eq:boundary_cond}
    \begin{aligned}
         \bar X'(0) = 0, \qquad \bar X(L) = 0,
    \end{aligned}
\end{equation}
for axial oscillations,  and 
\begin{equation} \label{eq:boundary_cond_y}
    \begin{aligned}
        & \bar Y(0) = \bar Y'(0) = \bar Y''(L) = 0,  \\
        & \bar Y'''(L) - (1+p)\, p\, q\, \frac{\pi^2}{L^2}\, \left[\bar Y'(L) + (1+p) \frac{f''(0)}{L}\, \bar Y(L) \right] = 0,
    \end{aligned}
\end{equation}   
for transverse ones.
Due to motion decoupling from Eqs. \eqref{eq:decoupled_osc}, $\omega_x$ and $\omega_y$ are independent from each other. Taking into account  the boundary conditions \eqref{eq:boundary_cond}, the integration of Eq. \eqref{eq:decoupled_osc}$_1$ provides the axial mode as
\beq
    \bar{X}(S)=\cos\!\left(\frac{\pi\,\hat\omega_x\,S}{L}\right),
\eeq
where
\beq
    \hat\omega_x=n-\frac{1}{2},\qquad n\in\mathbb{N},
\eeq
showing that the axial vibrations have always real circular frequencies $\omega_x$.
The transverse mode can be obtained from integration of Eq. \eqref{eq:decoupled_osc}$_2$  as 
\beq\label{eq:transvmode}
    \bar Y(S) = C_1\cos\!\left( \frac{\pi\lambda_1 S}{L}\right) + C_2\cos\!\left(\frac{\pi\lambda_2  S}{L}\right) +C_3\sin\!\left( \frac{\pi\lambda_1 S}{L}\right) +C_4\sin\!\left( \frac{\pi\lambda_2 S}{L} \right),
\eeq
where
\begin{equation}
    \lambda_{1,2} =\sqrt{\frac{1+p}{2}}\, \sqrt{-p\, q\pm\sqrt{(p\, q)^2+4 \hat\omega_y^2}},
\end{equation}
and $C_1$, $C_2$, $C_3$, and $C_4$ are the coefficients defining the transverse shape. The homogeneous linear problem in the coefficients $C_j$ ($j=1,...,4$), obtained by  imposing the  boundary conditions \eqref{eq:boundary_cond_y} to the transverse shape, Eq. \eqref{eq:transvmode},  represents  an eigenvalue problem where $\hat\omega_y^2$ is the eigenvalue and the coefficients $C_j$ define the eigenvector.
In the present problem, it is found that an infinite set of $\hat\omega_y^2$ values exists, where each value is related to a specific transverse mode, and that these values are real numbers. Therefore, it can be concluded that:
\begin{itemize}
    \item if the smallest eigenvalue  $\hat{\omega}_y^2$ is positive ($\min[\hat{\omega}_y^2]>0$), the  trivial configuration is stable because all the eigenmodes have real circular  eigenfrequency $\omega_y$;
    \item if the smallest eigenvalue  $\hat{\omega}_y^2$ is negative ($\min[\hat{\omega}_y^2]<0$), the  trivial configuration is unstable because at least one of the eigenmodes has an imaginary  circular  eigenfrequency $\omega_y$ (corresponding to a divergent oscillation);
    \item if one of the eigenvalues  $\hat{\omega}_y^2$ vanishes, the  trivial configuration is at the transition between the two previous cases for the specific eigenmode. This condition is the dynamic counterpart of the previously analysed bifurcation for quasi-static deformation.\footnote{The equation of motion \eqref{eq:decoupled_osc}$_2$, in the case when $\hat\omega_y^2=0$, is equivalent to the  bifurcation equation \eqref{eq:linearized_for_critical}, obtained for quasi-static deformation. These two just differ in the reference kinematic field, corresponding to  $\widehat Y(S)$ in dynamics and $\theta(S)$ for quasi-static deformation. These are related to each other through $\widehat Y'(S)=(1+p)\,\theta(S)$ under the small rotation assumption. 
        } 
\end{itemize}
As example, the smallest eigenvalue  $\hat{\omega}_y^2$  is reported as a function of the dimensionless load $p$ in  Fig. \ref{fig:eigenfreq},  for $f''(0)=-6$ and two different stiffness ratios, $q=6.5$ (left) and  $q=8.5$ (right), representative of sets $\mathcal{A}$ and $\mathcal{B}$, respectively. Load ranges realizing  stable (unstable) trivial configuration, and corresponding to positive (negative) smallest eigenvalue  $\hat{\omega}_y^2$, are highlighted as white (gray) regions. These plots confirm the dimensionless load ranges corresponding to stable and unstable straight configuration as expressed by Eq. (\ref{stableA}) for set $\mathcal{A}$  and by Eq.  (\ref{stableB}) for $\mathcal{B}$, and the associated restabilization and double-restabilization phenomena.

\section{Analytical solution for the nonlinear kinematics of the structure}\label{batistone}

As the rotation and deformation may reach large values in the post-critical response, the nonlinear problem defined by Eq. \eqref{eq:governing} is investigated as  two decoupled problems. The first is the underlying problem of the deformation of the extensible elastica, while the second is the satisfaction of the boundary condition for the right end that is constrained by the curved profile.

For the extensible elastica, 
the closed-form solution  for
a  cantilever beam with a force imposed at the free end provided in   \cite{batista2016closed} is exploited. In particular, this solution, 
based on Jacobi elliptic functions as an improvement over the elliptic integral approaches considered before \cite{humer2011elliptic, humer2013exact, magnusson2001behaviour,pfluger2013stabilitatsprobleme, stoker1968nonlinear}, is specialized  to impose  the constraint expressed by 
Eq. \eqref{eq:constraint}.

\subsection{General solution for a cantilever beam with a force imposed at its end}

The closed-form solution for a cantilever beam with  a concentrated load at the free end 
has been derived by Batista \cite{batista2016closed}. This solution is exploited here in the case of unshearable rod (parameter $\nu=1$ in \cite{batista2016closed}) and by considering that the  concentrated load is provided by  the   reaction force of the profile, having magnitude $R>0$ and inclination $\alpha\in[-\pi,\pi]$ as reported in Fig. \ref{fig:schematic}.  

Introducing the following non-dimensional parameters\footnote{It should be noted that here the symbol $\rho$ is used instead of $\omega$ used originally in \cite{batista2016closed} in order to not create confusion with the analogous symbols used to denote the angular frequencies in Sect.  \ref{sec:vibr}.}  
\begin{equation}
    \rho^2 = \frac{R\,L}{B},\quad \lambda = L\,\sqrt{\frac{K}{B}},\quad \hat S = \frac{S}{L},
\end{equation}
the rotation $\theta(S)$ and strain $\varepsilon(S)$ fields are given by
\begin{equation} \label{eq:psi2}
    \theta(S ) = 2\arcsin\left[ k \frac{\text{sn}(\tilde\rho\, \hat S + C,\tilde k)}{\sqrt{1+m^2\, \text{cn}^2(\tilde\rho\, \hat S+C,\tilde k)}} \right]-\alpha ,
    \qquad
    \varepsilon(S) = -\frac{\rho^2}{\lambda^2} \cos(\theta(S)+\alpha),
\end{equation}
where 
\begin{equation}
    m^2 = \dfrac{ \dfrac{\rho^2  k^2}{\lambda^2} }{1 - \dfrac{\rho^2(1-k^2)}{\lambda^2} },
\qquad
    \tilde\rho = \rho \sqrt{ 1 + \frac{\rho^2(2\, k^2 - 1)}{\lambda^2}  },
\qquad
    \tilde k^2 = \frac{k^2 + m^2}{1 + m^2}.
\end{equation}
The integration of the kinematic fields (\ref{eq:psi2})$_1$ through Eq. (\ref{positions}) provides the position of the rod{, in the rotated reference frame, } at the curvilinear coordinate $S$ as
\begin{equation} \label{eq:coords_analytical}
    \begin{array}{lll}
    \begin{array}{lll}
   x(S) = &
             \ds\frac{2 \tilde{\rho}}{\rho^2} \left\{ \left( \frac{\mbox{E}(\tilde k)}{\mbox{K}(\tilde k)} - \frac{1}{2} \right) \tilde\rho\, \hat S + \mbox{Z}(\tilde\rho\, \hat S + C, \tilde k) - \mbox{Z}(C,\tilde k) \right. \\[4mm]
             &\ds\left. - m^2 \left[ \frac{ \text{sn}(\tilde\rho \hat S+C, \tilde k)\, \text{cn}(\tilde\rho \hat S +C, \tilde k)\, \text{dn}(\tilde\rho \hat S+C, \tilde k) }{1 + m^2 \,\text{cn}^2(\tilde\rho \hat S+C, \tilde k)} - \frac{ \text{sn}(C, \tilde k)\, \text{cn}(C, \tilde k)\, \text{dn}(C, \tilde k) }{1 + m^2\, \text{cn}^2(C, \tilde k)} \right]   \right\}L ,
        \\[6mm]
        y(S) = &\ds \frac{2\, k\, \tilde \rho\, \sqrt{1+m^2}}{\rho^2} \left[ \frac{\text{cn}(C,\tilde k)}{1+m^2\,\text{cn}^2(C,\tilde k)} - \frac{\text{cn}(\tilde\rho \hat S + C,\tilde k)}{1+m^2\,\text{cn}^2(\tilde\rho \hat S + C,\tilde k)} \right]L,
    \end{array}
    \end{array}
\end{equation}
while  the rod's curvature $\kappa(S)=\theta'( S )$ follows as
\begin{equation}
    \kappa(S) = \frac{1}{L} \frac{ 2\, \tilde\rho\, k \sqrt{1+m^2}\, \text{cn}(\tilde\rho\, \hat S + C, \tilde k) }{ 1+m^2\, \text{cn}^2(\tilde\rho\, \hat S+C,\tilde k) }.
\end{equation} 
In the above equations, $\mbox{K}$ and $\mbox{E}$ are the complete elliptic integrals of the first and second kind, $\mbox{Z}$, $\text{sn}$, $\text{cn}$, $\text{dn}$ are Jacobi elliptic functions, and $C$ is the constant of integration that can be found by imposing the boundary condition of zero bending moment, equivalent to null curvature, at the free end, $\theta'(L)=\kappa(L)=0$.

From the boundary condition of null rotation at the  clamp, $\theta(0) = 0$, the following equation is obtained
\begin{equation}\label{terzaeq}
    k \frac{\text{sn}(C,\tilde k)}{\sqrt{1+m^2\,\text{cn}^2(C,\tilde k)}} - \sin\frac{\alpha}{2} = 0,
\end{equation}
to be solved for the value of $k$. The boundary condition at the right end $\kappa(L)=0$ 
yields a formula for $C$
\begin{equation}
    C = -\tilde\rho + (2n-1)\, \mbox{K}(\tilde k),\qquad n\in\mathbb{N},
\end{equation}
where $n$ is the mode number. Once $k$ is determined, the coordinates of the curve in the rotated reference frame $x-y$ can be evaluated from  Eq. \eqref{eq:coords_analytical}, and those  in the reference frame $X-Y$, moving with the sliding clamp, follows as
\begin{equation}
    X(S) = x(S)\,\cos\alpha + y(S)\,\sin\alpha,\quad Y(S) = -x(S)\,\sin\alpha + y(S)\,\cos\alpha.
\end{equation}
Finally, recalling from Eq. (\ref{eq:zozza}) that 
\begin{equation}
    X(L) = d_X, \qquad Y(L)  = d_Y,
\end{equation}
these two quantities can be expressed as a respective function, $\mathcal{G}_X$ and $\mathcal{G}_Y$ (for conciseness not explicitly reported here), depending on the angle $\alpha$ and the reaction magnitude $R$, 
\begin{equation}\label{GiGi}
    d_X= \mathcal{G}_X(\alpha, R), \quad d_Y= \mathcal{G}_Y(\alpha, R).
\end{equation}
Because $\kappa(L)=0$, the parameter $k$ can be related with the rotation of the beam at the free end $\theta_L=\theta(L)$, by evaluating \eqref{eq:psi2} at $S=L$, through the following formula
\begin{equation}
    k = %\sin\frac{\psi_L}{2} =
    \sin\frac{\theta_L + \alpha}{2}.
\end{equation}

\subsection{Imposition of the curved profile constraint}

The general solution for a cantilever beam recalled in the previous section is now considered as having unknown   reaction force $R$ and  angle $\alpha$. 
Because the load $P$ is imposed at the sliding clamp, 
equilibrium implies the following relation between the angle $\alpha\in[-\pi,\pi]$ and the positive reaction force $R$
\begin{equation}
    \label{PRalfa}P = R\,\cos\alpha .
\end{equation}
Due to the absence of  friction, the reaction force $R$ is perpendicular to the constraint profile, relating $\alpha$ with the vertical deflection of the free end of the cantilever beam $d_Y$ as\footnote{The definition \eqref{eq:alpha_ref} for the angle $\alpha$ is introduced to ensure positiveness of the profile reaction,  $R>0$, independently from the sign of the applied load $P$. Moreover, the restriction $R>0$  is also needed to exploit the solution obtained in \cite{batista2016closed}.}
\begin{equation} 
\label{eq:alpha_ref}
    \alpha = \arctan\!\left(f'\left(\dfrac{d_Y}{L}\right)\right) + \frac{\pi}{2} [ 1-\text{sgn}(P) ] .
\end{equation}

Considering Eq. (\ref{PRalfa}), relation (\ref{GiGi}) can be rewritten as
\begin{equation} \label{eq:Y1_fixedpoint}
    d_X =\mathcal{G}_X\!\left(\alpha, \frac{P}{\cos\alpha}\right)\!,\qquad
    d_Y =\mathcal{G}_Y\!\left(\alpha, \frac{P}{\cos\alpha}\right)\!,
\end{equation}
and, since the angle $\alpha$ is related to $d_Y$ through 
Eq. \eqref{eq:alpha_ref}, Eq. \eqref{eq:Y1_fixedpoint}$_2$ provides a fixed-point recursive relation for the value of $d_Y$.

The nonlinear equilibrium configurations can be obtained by solving a  reverse problem, where the displacement $\Delta$ is imposed while the load $P$ is  unknown. In this way, Eqs. \eqref{eq:constraint},   \eqref{eq:Y1_fixedpoint}, and \eqref{terzaeq} provide a system of three nonlinear equations in the three unknown parameters $P$, $\theta_L$, and $d_Y$ to be solved for a given $\Delta$ value
\begin{equation}\label{nonlinsys}
\begin{cases}
    \begin{aligned}
        & \mathcal{G}_X\!\left(\alpha,\frac{P}{\cos\alpha}\right) - L\, \left[1 + f\!\left(\frac{d_Y}{L}\right)\right] - \Delta = 0, \\[1em]
        & \mathcal{G}_Y\!\left(\alpha,\frac{P}{\cos\alpha}\right) - d_Y = 0, \\[1em]
        & k \frac{\text{sn}(C,\tilde k)}{\sqrt{1+m^2\,\text{cn}^2(C,\tilde k)}} - \sin\frac{\alpha}{2} = 0.
    \end{aligned}
\end{cases}
\end{equation}

Due to the nonlinearity of system (\ref{nonlinsys}), none, one, or multiple non-trivial equilibrium configurations can be obtained for a given value of  the end displacement $\Delta$.

\subsection{Symmetric profile}

Reference is made to a 
parabolic 
constraint profile (symmetric about the $\widehat{X}$--axis) and described by 
\beq
    f(\xi) = 1 + \frac{f''(0)}{2}\, \xi^2,
\eeq
where $\xi = \widehat{Y}/L$. 

\begin{figure}
    \centering
        \includegraphics[width=150mm]{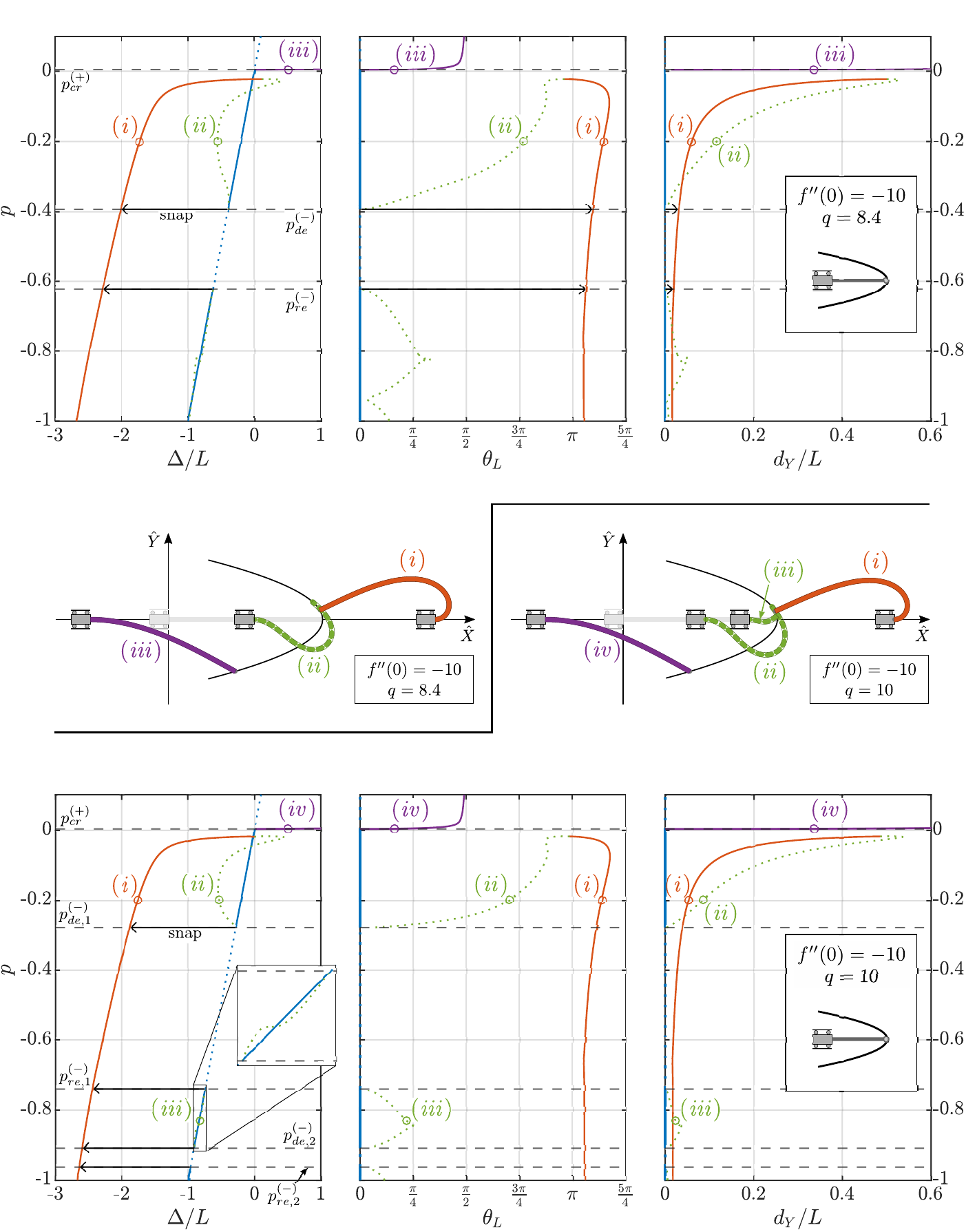}
    \caption{\label{fig:fd_parabolic_10}
   \footnotesize Post-critical behaviour in terms of dimensionless force $p$ versus different measures of structural deformation: clamp displacement $\Delta$, right end rotation $\theta_L$ and vertical displacement $d_Y$.
    $f''(0)=-10$ and $q=8.4$ (upper part) and $q=10$ (lower part). Some stable and unstable  deformed configurations are shown for specific  non-trivial states in the central part.
    }
\end{figure}

The equilibrium paths in terms of load $p$ as a function of the clamp displacement $\Delta/L$, rotation $\theta_L$, and vertical coordinate $d_Y/L$
of the end of the rod 
are shown in Fig. \ref{fig:fd_parabolic_10} for  $f''(0)=-10$. Two different values of $q$ are considered, $q=8.4$ 
and $q=10$,  corresponding to a single and a double restabilization, respectively. Stable and unstable configurations are distinguished through continuous and dotted lines, respectively.\footnote{Stability character of the non-trivial configurations and snap motion towards them have been assessed through the numerical solution of the nonlinear dynamic equations \eqref{eq:vibr_eom} and \eqref{eq:vibr_bc}.} Restricting attention to non-trivial configurations, the force-displacement diagrams reveal one stable non-trivial  configuration (highlighted with red lines) and, as a result, the system is bistable. 
Because  none of the non-trivial bifurcation paths (highlighted with green dotted lines) is stable in compression, the loss of stability at these bifurcation points realizes  a \lq snap' motion towards the non-trivial stable configuration, as shown by the arrows in Fig.  \ref{fig:fd_parabolic_10}. 
Moreover, since $f''(0)$ is negative, a tensile bifurcation exists. The non-trivial path (highlighted with purple lines) at the tensile bifurcation is found to be stable.

%%%%%%%%%%%%%%%%%%%%%%%%%%%%%%
%%%%%%%%%%%%%%%%%%%%%%%%%%%%%%

\subsection{Skew-symmetric profile with discontinuous curvature at the origin}
\begin{figure}[t]
    \centering
    \includegraphics[width=75mm]{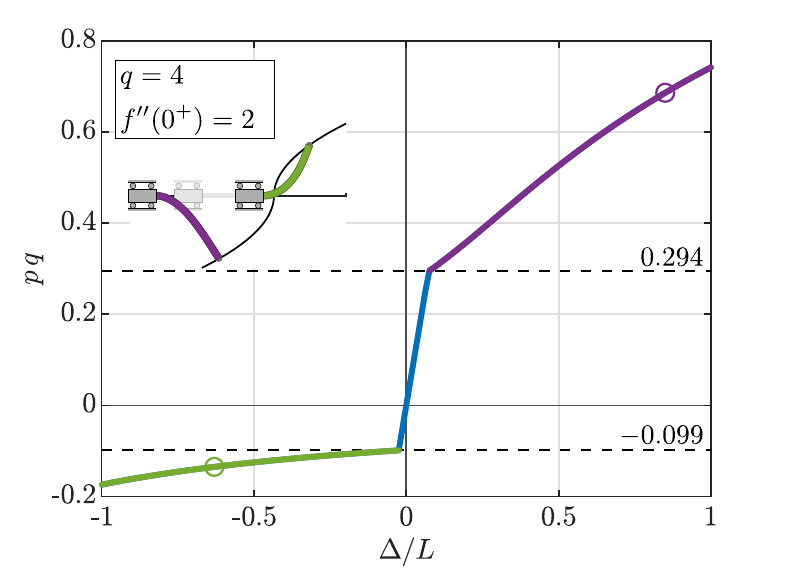}
    \includegraphics[width=75mm]{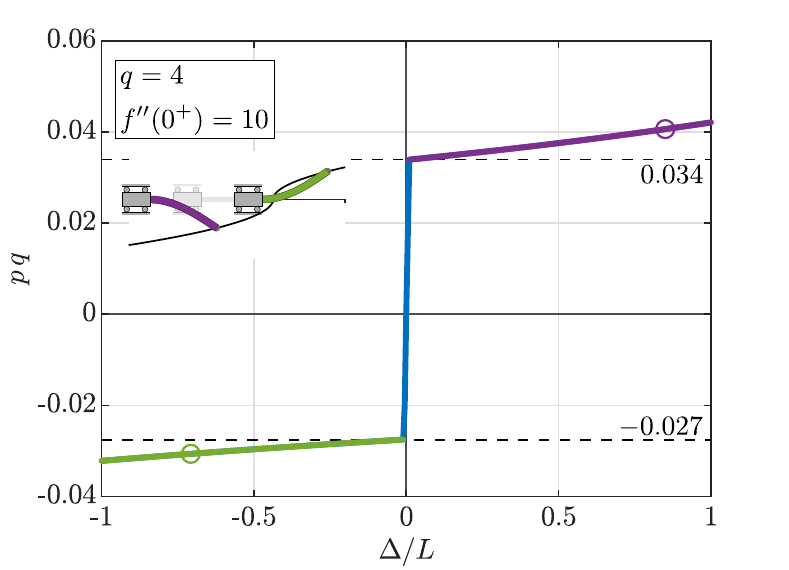}
        \includegraphics[width=75mm]{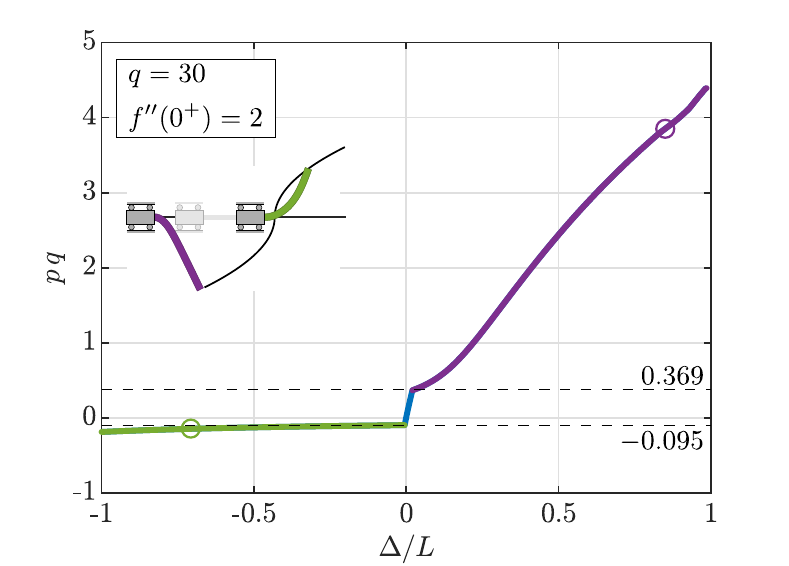}
    \includegraphics[width=75mm]{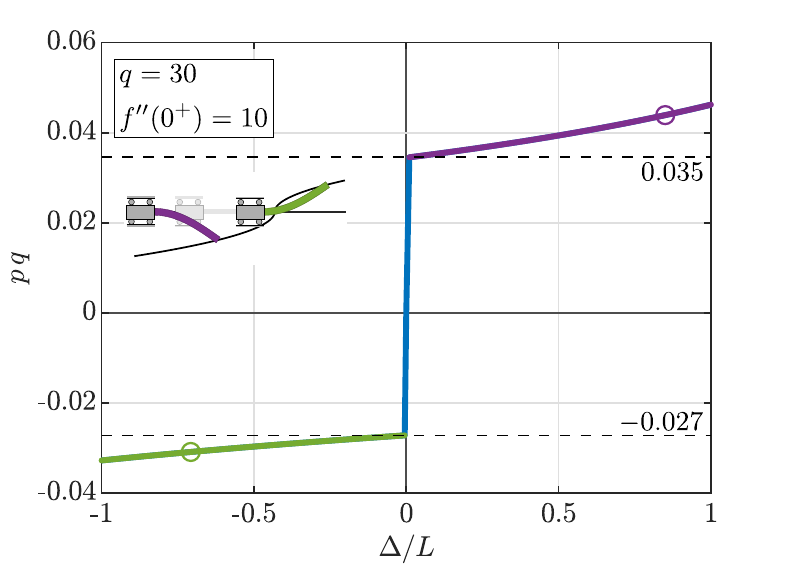}
    \caption{\label{fig:fd_parabolic_simple} \footnotesize Force  ($\,p q=P L^2/(\pi^2 B)$) -- displacement ($\Delta/L$) diagram for a structure with  $q=4,\,30$ and skew-symmetric profile  defined by Eq. \eqref{shape} with $f''\left(0^+\right)=2$ (left) and $f''\left(0^+\right)=10$ (right). The deformed shapes shown in the inset correspond to the $p q$--$\Delta/L$ pairs highlighted with a circle (with corresponding colour) on the non-trivial path. }
\end{figure}

A profile, skew-symmetric  about the $\widehat{X}$--axis, is considered  as
\begin{equation}\label{shape}
    f(\xi) = 1 + \frac{f''(0^+)}{2}\, \xi\,\left|\xi\right|,
\end{equation}
displaying a discontinuity in the profile curvature at the origin,
\beq
f''\left(0^-\right)=-f''\left(0^+\right).
\eeq
For this system, attention is restricted only to the continuous stable path in terms of  $p\,q$--$\Delta/L$, reported in  Fig. \ref{fig:fd_parabolic_simple} for  $q=\left\{4,30\right\}$ and $f''\left(0^+\right)=\{2,\,10\}$. It can be noted that, although the profile shape is skew-symmetric about the $\widehat{X}$--axis, the force-displacement curve does not evidence any symmetry, except along the trivial branch before bifurcation. In fact, the post-buckling behavior of the system depends not only on the constraint shape but also on the signs of the profile angle $f'(\xi)$ and curvature $f''(\xi)$.

%%%%%%%%%%%%%%%%%%%%%%%%%%%%%%
%%%%%%%%%%%%%%%%%%%%%%%%%%%%%%

\section{Optimization of the  profile shape for a design force-displacement curve}
\label{sec:optim}

The developed theoretical framework  
is  exploited here 
to design a device displaying a prescribed   force-displacement $p-\Delta/L$ curve.
The structure displays a  linear elastic range, limited within the tensile/compressive bifurcations, $p=\Delta/L \in\left[p_{cr}^{(-)}, \, p_{cr}^{(+)}\right]$, defined by Eq. (\ref{ppcrcr}). 
The design force-displacement curve is assumed to be characterized as 
\begin{equation} \label{eq:generic}
    p\!\left(\dfrac{\Delta}{L}\right) = 
    \left\{
    \begin{array}{ll}
        \mathsf{p}^{(-)}\!\left(\dfrac{\Delta}{L}\right),  & \dfrac{\Delta}{L} <p_{cr}^{(-)} ,        \\[4mm]
        \dfrac{ \Delta}{L}, & \dfrac{\Delta}{L}\in\left[p_{cr}^{(-)} ,\,p_{cr}^{(+)} \right], \\[4mm]
        \mathsf{p}^{(+)}\!\left(\dfrac{\Delta}{L}\right),  & \dfrac{\Delta}{L} > p_{cr}^{(+)},
        \end{array}
        \right.
\end{equation}
where $\mathsf{p}^{(+)}(\Delta/L)$ and $\mathsf{p}^{(-)}(\Delta/L)$  define a target behaviour to be displayed after  bifurcation, tensile and compressive, respectively, and satisfying the continuity condition
\beq
 \mathsf{p}^{(\pm)}\!\left( 
p_{cr}^{(\pm)}
 \right)=\,p_{cr}^{(\pm)}.
  \eeq
  
Considering that the two critical values are independent of each other, a discontinuous profile curvature at the origin is needed to  control independently these two thresholds as related to the tensile and compressive loading branches of the structure. As previously shown, the stiffness ratio $q$ and two profile curvatures at the origin $f''(0^-)$ and $f''(0^+)$ have to lie within the ranges provided by Eq. \eqref{ranges}.

  The post-buckling displacement $\Delta$ is discretized in the two  sets $\Delta_i^{(\pm)}$,
  characterized by $N^{(\pm)}$  discretization points ($i=0,...,N^{(\pm)}$), satisfying
    \beq
  \Delta_0^{(\pm)}=p_{cr}^{\pm}\, L,\qquad
  \left|\Delta_{i+1}^{(\pm)}\right|>\left|\Delta_{i}^{(\pm)}\right|,\qquad
  \pm \Delta_i^{(\pm)}>0.
  \eeq
  Considering the profile as the union of two parts differing in the sign of $Y$, the dimensionless coordinate $Y/L$ is discretized through  the unknown points $\xi_i^{(\pm)}$, corresponding to $\Delta_i^{(\pm)}$, where
  \beq
  \xi_0^{(\pm)}=0^{\pm},\qquad
  \left|\xi_{i+1}^{(\pm)}\right|>\left|\xi_{i}^{(\pm)}\right|,\qquad
  \pm \xi_i^{(\pm)}>0.
  \eeq
  The profile shape 
  is discretized as 
  \beq
  f(\xi)=1+\dfrac{1}{2}\sum_{j=0}^{i-1}h_{j+1}^{(\pm)} \left(\xi^{(\pm)}_{j+1}-\xi^{(\pm)}_j\right)^2+\dfrac{1}{2}h_{i+1}^{(\pm)}\left(\xi-\xi^{(\pm)}_i\right)^2, \qquad \xi\in\left[\xi^{(\pm)}_i,\,\xi_{i+1}^{(\pm)}\right],
  \eeq
  where $h_{i+1}^{(\pm)}$ is the piecewise constant profile curvature value,
  \beq
  f''(\xi)=h_{i+1}^{(\pm)}, \qquad  \xi\in\left(\xi^{(\pm)}_i,\,\xi_{i+1}^{(\pm)}\right),
  \eeq

In order to achieve the target force-displacement curve, a fitting loop is devised. 
The algorithm initiates with $i=0$ and is repeated for increasing values of $i$. After selecting the stiffness ratio $q$ of the rod, the  generic unknown point $\xi_{i+1}^{(\pm)}$ and corresponding profile curvature value $h_{i+1}^{(\pm)}$ can be found through a  solution of the nonlinear system (\ref{nonlinsys}) by imposing $\Delta_{i+1}^{(\pm)}$ and prescribing the force $p$ in agreement with the target response, Eq. (\ref{eq:generic}), as
\begin{equation} \label{eq:fit_cond}
    p\!\left(\dfrac{\Delta_{i+1}^{(\pm)}}{L};\, \xi_{i+1}^{(\pm)},\, h_{i+1}^{(\pm)}\right) =\, \mathsf{p}^{(\pm)}\!\left(\dfrac{\Delta_{i+1}^{(\pm)}}{L}\right)\!.
\end{equation}

It is worth noting that the profile optimization can be unsuccessful at a specific value of $\Delta$ for the following issues. 
\begin{enumerate}[label={Issue \arabic* --},leftmargin=*]
    \item A too  stiff response is requested. Indeed, the target force-displacement curve $\mathsf{p}^{(\pm)}\left(\Delta/L\right)$ has its first derivative bounded from above by the maximum post-critical stiffness ratio $k_{\text{max}}\in(0,1)$ as\footnote{The profile  can be designed for realizing a post-bifurcation response with incremental negative stiffness,
\beq
\dfrac{\mbox{d}\mathsf{p}^{(\pm)}\left(\Delta/L\right)}{\mbox{d}\left(\Delta/L\right)}<0,
\eeq
a concept under research for applications in vibration isolation and seismic engineering; see e.g. \cite{Antoniadis2018KDampingAS,liu2013characteristics,sapountzakis2017kdamper,zheng2016design}. }
\beq\label{kmaxxxx}
\dfrac{\mbox{d}\mathsf{p}^{(\pm)}\!\left(\Delta/L\right)}{\mbox{d}\!\left(\Delta/L\right)}<k_{\text{max}}\!\left(q,\widehat{Y}\right),
\eeq
because a unit value would correspond to the stiffest behaviour given by the purely axial state,  which is instead weakened  in the post-critical  region by flexure. It is also observed that high values of $q$ and $\widehat{Y}$ define small    maximum post-critical stiffness ratio $k_{\text{max}}$;
\item The elastica is unstable. Although a profile shape is found, the involved equilibrium configuration of the rod may become unstable. In particular, the stable/unstable response is affected by the prescribed boundary condition, namely, it may change whether displacement $\Delta$ or load $p$ are imposed;
\item A force reversal {\cite{bigoni2012effects,bosi2015self,BOSI201583}} is excluded. Because  the profile shape is described via the horizontal coordinate $\widehat X$ through the function $f\!\left(\widehat Y/L\right)$, Eq. (\ref{eq:constraintprofile}), the present formulation  cannot be exploited to realize force reversals ($p$ changing sign at a given end displacement  $\Delta\neq0$). Therefore,  the target response is subject to the following constraint
\beq\label{forcerev}
\pm\mathsf{p}^{(\pm)}>0.
\eeq
 In the limit condition given by $p=0$, the optimization procedure fails  providing  $f'\rightarrow\infty$, because the only possible  equilibrium configuration is realized for  a reaction inclination angle $\alpha=\pm \pi/2$.
\end{enumerate}

While the first two issues above are inherent to the considered model and cannot be avoided, the latter restriction  can be  overcome through a parametric  planar description of the profile curve, as for example describing the profile as $\hat X(\sigma)$ and $\hat Y(\sigma)$  through a  parameter $\sigma$.

In the following, examples of application of the optimization algorithm are presented by considering a uniform spacing  $\Delta_{i+1}^{(\pm)}-\Delta_{i}^{(\pm)}=\left(\Delta_N^{(\pm)}-\Delta_{0}^{(\pm)}\right)/N^{(\pm)}$ in the discretization, with $N^{(\pm)}=120$, $\Delta_N^{(\pm)}=2L$.

The following  examples are also complemented by movies (available as electronic supplementary material) showing the theoretical behaviour of the structure  optimized to display specific force-displacement curves.

\subsection{The design of an elastic force-limiting device} \label{sec:opt_fl}

In an ideal force-limiting device, 
the force $P$ is expected  to be initially linear with the displacement  $\Delta$, up to a threshold value at which the bifurcation occurs. Subsequently,  
the force is requested to remain constant (or slightly raise in its absolute value) at a further increase in the displacement magnitude, up to a point where the structure locks and the stroke of the device is attained. 
In practice however, a small increase in the force magnitude after buckling is desirable, to facilitate the return of the device to its initial configuration at unloading and to avoid snap motion due to negative stiffness. By excluding for the moment the final locking part, this behaviour can be described in mathematical terms as a symmetric bilinear  force-displacement curve, namely, Eq. \eqref{eq:generic} with $\mathsf{p}^{(\pm)}$ given as
\begin{equation} \label{eq:bilinear}
    \mathsf{p}^{(\pm)}\!\left(\dfrac{\Delta}{L};   p_{cr},\, r\right) = \pm \, p_{cr}
         + r\,\left(\dfrac{\Delta}{L}\mp p_{cr}\right),
         \end{equation}
where $r$ is the bilinear stiffness ratio, relating the incremental constant stiffness in  the post-buckling regime to that characterizing the structure before bifurcation (therefore enforced by Eq. (\ref{kmaxxxx}) to $r<k_{\mbox{max}}<1$), and  $p_{cr}=p_{cr}^{(+)}=-p_{cr}^{(-)}$.

The constraint's profile, which realizes a certain designed force-displacement behaviour (defined by $p_{cr}$ and $r$), is not unique, rather, various shapes of the profile, realizing  the target post-buckling behaviour $\mathsf{p}^{(\pm)}$, can be found using the algorithm described in the previous section as a function of $q$. 

\begin{figure}[ht!]
    \centering 
    \includegraphics[width=\textwidth]{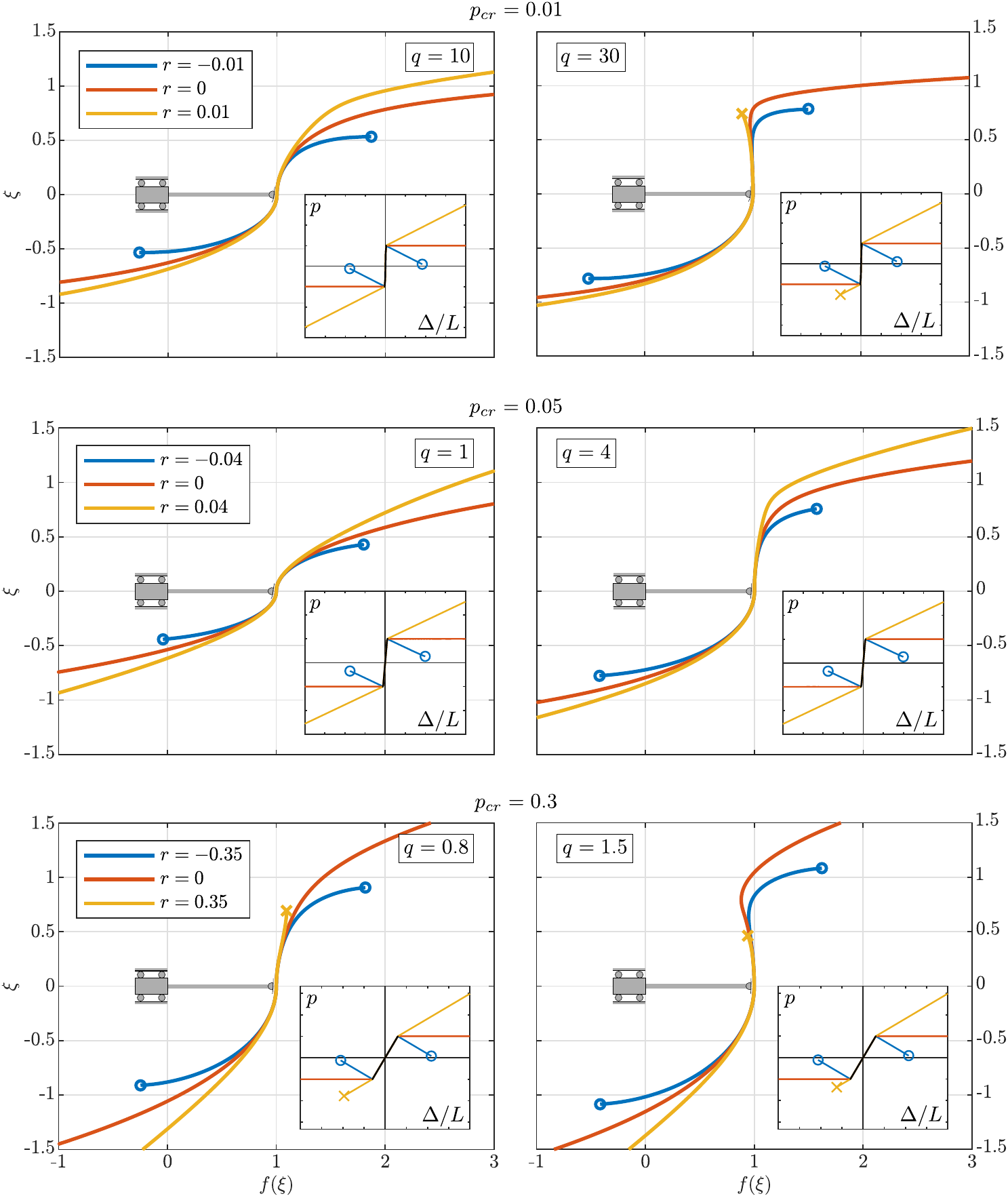}
    \caption{\label{fig:optimized_20} \footnotesize Profile's shapes realizing a symmetric bilinear force-displacement $p-\Delta$ curve (shown in the insets) with { $p_{cr}=\left\{0.01, 0.05, 0.3\right\}$} (from top to the bottom)
    for three different values of the bilinear stiffness ratio $r=\{-0.01,0,0.01\}$ (top), $r=\{-0.04,0,0.04\}$ (middle) and $r=\{-0.35,0,0.35\}$ (bottom).
    % $r=\{-0.04,0,0.04\}$ (left) and $r=\{-0.35,0,0.35\}$ (right). 
    Crosses and circles denote the points where the design $p-\Delta$ fails, putting out-of-service the device. }
\end{figure}

A number of constraint's profile shapes are shown in Fig. \ref{fig:optimized_20} for $p_{cr}=\left\{0.01, 0.05, 0.3\right\}$ (from top to the bottom).
% $p_{cr}=0.05$ (left) and $p_{cr}=0.3$  (right). 
The same force-displacement behaviour (reported in the insets) is obtained by changing $q$, assuming 2 pairs of values reported in each diagram.
Profile shapes with different colours correspond to different values of bilinear stiffness ratio  $r=\{-0.01,0,0.01\}$ (top), $r=\{-0.04,0,0.04\}$ (middle) and $r=\{-0.35,0,0.35\}$ (bottom).
% $r=\{-0.04,0,0.04\}$ (on the left)  and $r=\{-0.35,0,0.35\}$ (on the right). 

%%%%%%%%%%%%%%%%%%%%%%%%%%%%%%%%%%%%%%%%
%%%%%%%%%%%%%%%%%%%%%%%%%%%%%%%%%%%%%%%%

\begin{figure}[ht!]
    \centering 
    \includegraphics[width=\textwidth]{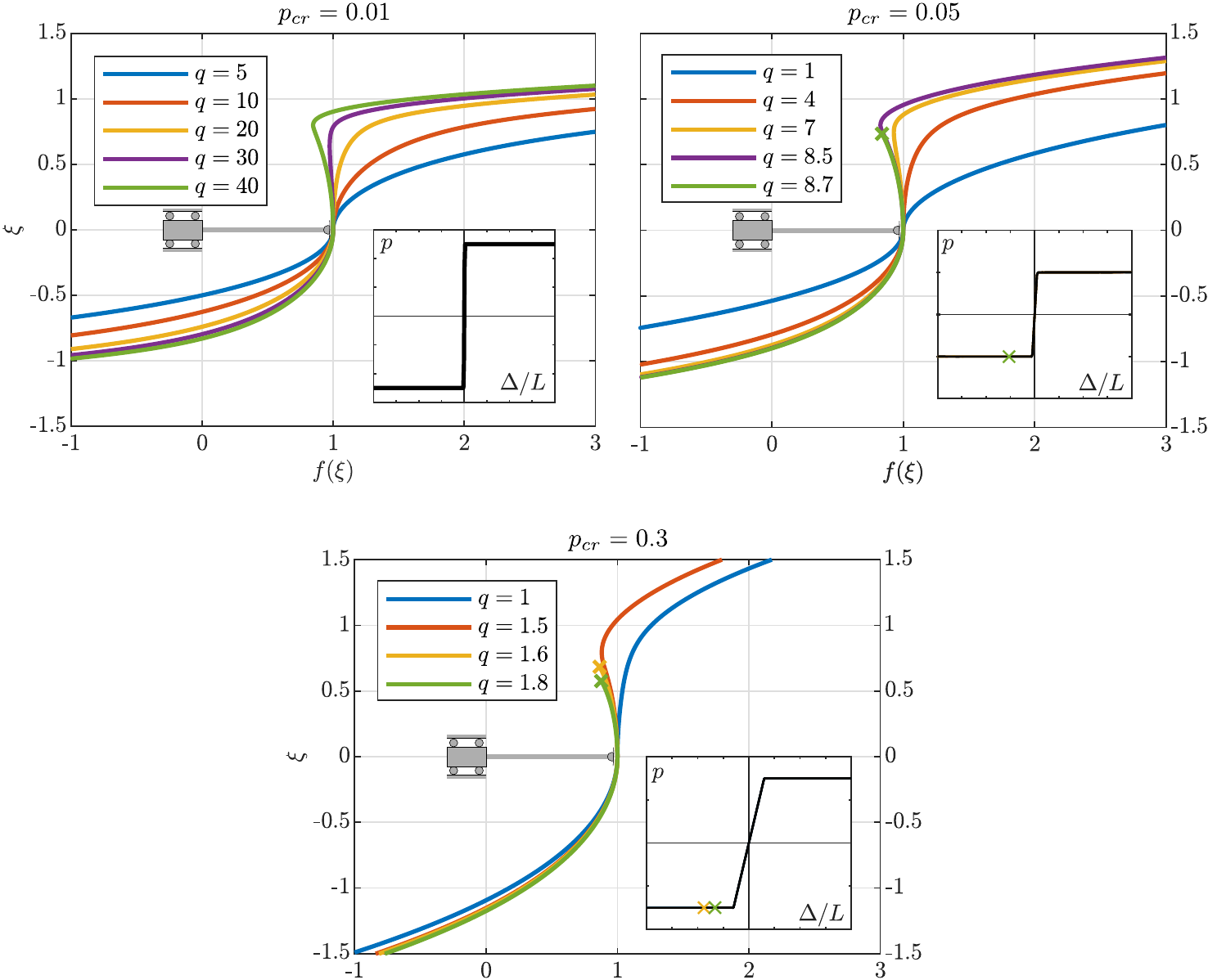}
    \caption{\label{fig:optimized_varq} \footnotesize Profile's shapes realizing a symmetric bilinear force-displacement $p-\Delta$ curve with null bilinear stiffness ratio (shown in the insets), $r=0$,  for $p_{cr}=0.05$ (top left), $p_{cr}=0.05$ (top right) and $p_{cr}=0.3$ (bottom). 
    % $p_{cr}=0.05$ (left) and $p_{cr}=0.3$ (right). 
    Crosses denote the points where the design $p-\Delta$ fails, putting out-of-service the device.  Note that small values of $q$ extend the range of displacement for which the designed response is displayed. }
\end{figure}

Restricting attention to a null bilinear stiffness ratio, $r=0$, the profile shapes are shown in Fig. \ref{fig:optimized_varq}
for $p_{cr}=0.01$ (top left), $p_{cr}=0.05$ (top right), and $p_{cr}=0.3$ (bottom),
% $p_{cr}=0.05$ (left) and $p_{cr}=0.3$ (right), 
for different values of the stiffness ratio $q$. The force-displacement diagrams are reported in the insets. 

In both Figs.  \ref{fig:optimized_20} and \ref{fig:optimized_varq}, some of the profile shapes and the corresponding load-displacement curves are ended with a marker (a cross or a circle). This marker shows the point where the optimization algorithm fails to converge because the target stiffness is too high (Issue 1, marked with a cross), or  because a force reversal is found (Issue 3, marked with a circle). It should be noted that the blue curves in Fig. \ref{fig:optimized_20} show negative stiffness and therefore are unstable for imposed load $p$.

Also, small values of $q$ provide smoother profiles than those obtained for large $q$. Moreover, small values of $q$ lead to a large stroke, defined as the greatest displacement $\Delta$ reached with the device, see Fig. \ref{fig:optimized_varq}, showing that rod's extensibility facilitates the profile optimization. It is noted that small values of $q$ can be obtained by using elements such as those described in Sect. \ref{premessa}. On the other hand, rods that are inextensible ($q\to\infty$) could be used to achieve the same effect, although inextensibility reduces the performance of the device. 

%%%%%%%%%%%%%%%%%%%%%%%%%%%%
%%%%%%%%%%%%%%%%%%%%%%%%%%%%

For the practical realization of a force-limiter, a secondary increase in the stiffness has to be introduced in order to achieve a smooth transition to a maximum allowable displacement $\Delta_{\mbox{max}}$. This secondary increase in the stiffness can be achieved by modifying the final part of the profile shape. An example for $p_{cr}=0.01$,   $q=10$, $r=0$, and locking starting at $\Delta/L=\pm0.73$
     is shown in Fig. \ref{fig:intro}, along with   the  profile shape and the deformed configuration at  $\Delta/L=\pm\{0.1, 0.2, 0.8\}$.

\subsection{Profile shape for complex force-displacement curves} \label{sec:opt_pattern}

To show how far the present framework can be exploited to design constraint's profiles 
imposing highly complicated load-displacement curves, the  present analysis is concluded by finding the profile shapes for a {\it sinusoidal}  post-critical response (with constant average load),
\begin{equation} \label{eq:sinus}
    \mathsf{p}^{(\pm)}\!\left(\dfrac{\Delta}{L};  p_{cr},\, a,\, b\, \right) =\, \pm \, p_{cr}
         + a \sin\left[2\pi\,b\left(\dfrac{\Delta}{L}\mp p_{cr}\right)\right]
\end{equation}
and  for a {\it triangular}  post-critical response (with non-constant average load),
\begin{equation} \label{eq:trian}
    \mathsf{p}^{(\pm)}\!\left(\dfrac{\Delta}{L};   p_{cr},\, r_1,\, r_2,\, c \right) =\, \pm \, p_{cr}
         + r_1\left(\dfrac{\Delta}{L}\mp p_{cr}\right)
         \mp r_2
         \,\left|c\left( \dfrac{\Delta}{L}\mp p_{cr}\right)
         -\left\lfloor
        c\left( \dfrac{\Delta}{L}\mp p_{cr}\right)\! +\dfrac{1}{2}\right\rfloor
         \right|,
         \end{equation}
         with the symbol $\left \lfloor{\cdot}\right \rfloor$ standing for the integer part of the relevant argument. 
Both the above design post-critical  responses define a multistable element, where more than one stable displacement $\Delta$ exist at a given load $p$.

\begin{figure}[ht!]
    \centering
    \includegraphics[width=\textwidth]{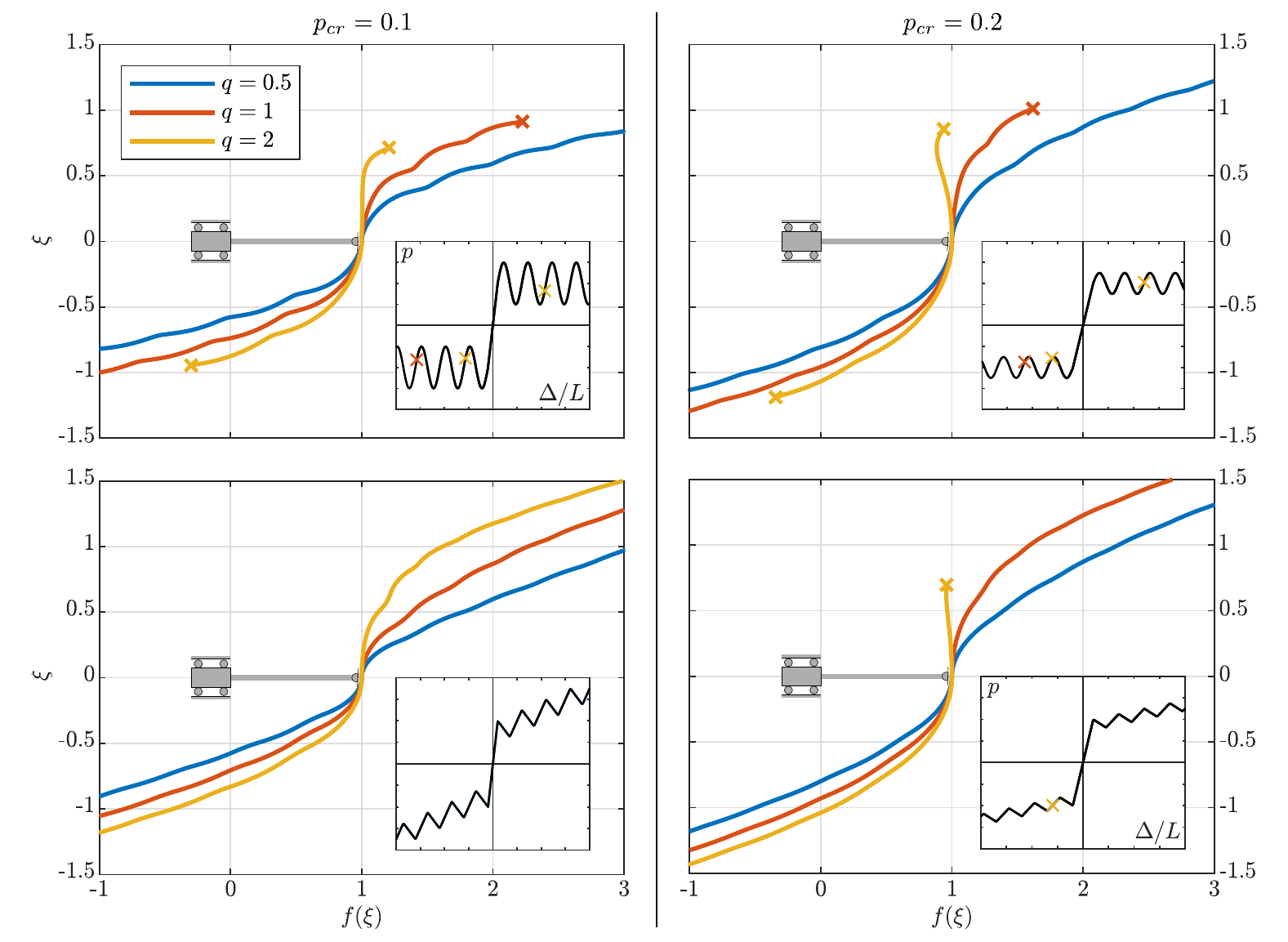}
    \caption{\label{fig:fd_other} \footnotesize Constraint's profile shapes optimized for realizing (above)  a sinusoidal
    force-displacement response, Eq. (\ref{eq:sinus}), with  $a=0.05$ and $b=2$ and (bottom) a  triangular response, Eq. (\ref{eq:trian}), with $r_1=0.05$, $r_2=0.1$, and $c=2$.
Critical load is $p_{cr}=0.1$ (left) and $p_{cr}=0.2$ (right). Different shapes are reported for different values of the stiffness ratio $q$. }
\end{figure}

The profile shape for the sinusoidal response (\ref{eq:sinus}) with  $a=0.05$ and $b=2$ is reported in Fig. \ref{fig:fd_other} (upper row) while that for the triangular one (\ref{eq:trian}) with $r_1=0.05$, $r_2=0.1$, and $c=2$  in Fig. \ref{fig:fd_other} (lower row).
Two values of critical load are prescribed, $p_{cr}=0.1$ (left) and $p_{cr}=0.2$ (right), while the stiffness ratio assumes three values, $q=\{0.5,1,2\}$. 
Similarly to the bilinear post-critical response, also for these other two cases, it  is evident that the profile shape becomes smoother and can be defined for greater $\Delta$ and for small values of $q$.

%%%%%%%%%%%%%%%%%%%%%%%%%%%%%%%%%%%%%%%%%%%%%%         
%%%%%%%%%%%%%%%%%%%%%%%%%%%%%%%%%%%%%%%%%%%%%

\section{Conclusions}
The extensible elastica has been analytically solved with an end constrained to move along a curved frictionless profile (with or without a discontinuous curvature value in correspondence of the straight configuration). The solution for the full non-linear structural response has  revealed the following features.
\begin{itemize}
    \item The values of profile curvature at the origin $f''(0)$ and of axial/flexural  stiffness ratio $q$ have a significant effect on the existence, number, and value of the bifurcation loads, which may be compressive or tensile;
    \item When bifurcation in compression occurs, the system may display a  single or a double restabilization of the straight configuration (occurring at large compression, depending on  $f''(0)$ and $q$).
\end{itemize}

An  optimization algorithm has been proposed for the  design of the profile shape, to obtain a prescribed 
post-critical response. The optimization algorithm 
was positively tested to realize a large variety of load-displacement curves (bilinear, sinusoidal, triangular), useful for applications as   force-limiter or other passive mechanisms.

\section*{Acknowledgements}
PK and FDC gratefully acknowledge the financial support from the European Union's Horizon 2020 research and innovation programme under the Marie Sklodowska-Curie grant agreement {`INSPIRE} - Innovative ground interface concepts for structure protection' {PITN-GA-2019-813424-INSPIRE}. 
DB gratefully acknowledges financial support from ERC-ADG-2021-101052956-BEYOND. Support 
from the Italian Ministry of Education,
University and Research (MIUR) in the frame of the
`Departments of Excellence' grant L. 232/2016 is acknowledged. This work has been developed under the auspices of INDAM-GNFM.

PK would like to thank Samuele Infanti (FIP MEC, Padova, Italy) for the useful discussions on technological aspects of the device design and realization. 

DB and FDC acknowledge the in-depth discussions and fruitful collaboration through these years with Natasha and Sasha. They look forward to many more years of enjoying together stimulating scientific interaction and friendship.

\section*{References}
\printbibliography[heading=none]

\appendix
%%%%%%%%%%%%%%%%%%%%%%%%%%%%%%%%%%%%%%%%%%%%%%%%
\renewcommand{\theequation}{{\thesection.}\arabic{equation}}
\renewcommand\thefigure{\thesection.\arabic{figure}}   
%%%%%%%%%%%%%%%%%%%%%%%%%%%%%%%%%%%%%%%%%%%%%%%%%%%%%%%%%%%%%%%%%%%%%%%

\section{Examples of one-dimensional structures displaying small values of stiffness ratio $q$} \label{ap:springs}
\setcounter{equation}{0}
\setcounter{figure}{0}

To further substantiate the examples introduced in Sect. \ref{premessa}, a  stiffness ratio evaluation is provided. The evaluation is based on the shear stiffness $K_s$, axial stiffness $K_a$ (corresponding to $K$ in the main text), and bending stiffness $B$ of the Reissner beam equivalent to a helical spring \cite{kobelev2021durability}
\begin{align}
    K_s^{\text{spring}} = \frac{8 E I_r L}{\pi n_a D^3} ,\qquad
    K_a^{\text{spring}} = \frac{4 G I_T L}{\pi n_a D^3} ,\qquad
    B^{\text{spring}}=   \frac{2 E I L}{\pi n_a D \left(1+\frac{E I}{G I_T}\right)},
\end{align}
where $E$ is the Young's modulus, $G$ the shear modulus, $I$ is the moment of inertia of the wire cross-section with respect to the radius of the spring coil passing through the centre of the cross-section, $I_r$ is the moment of inertia of the wire cross-section with respect to the axis perpendicular to the radius passing through the centre of the cross-section, $I_T$ is the torsion constant, $L$ is the length of the spring, $D$ is the coil diameter, and $n_a$ is the number of (active) coils.

The dimensionless axial stiffness parameter $q_a$  (corresponding to $q$ in our manuscript) and shear stiffness parameter $q_s$
are introduced  as
\begin{align}
    q_a = \frac{K_a L^2}{\pi^2 B} ,\qquad      q_s = \frac{K_s L^2}{\pi^2 B},
\end{align}
which reduce for a helical spring to
\begin{align}
    q_a^{\text{spring}}=  \frac{2}{\pi^2} \left(1+\frac{G I_T}{E I}\right) \left(\frac{L}{D}\right)^2,\qquad  q_s^{\text{spring}} = \frac{2 E I_r}{G I_T} q_a^{\text{spring}}.
\end{align}
Interestingly, the ratio between the two stiffness ratios is given by
\begin{equation}
    \frac{q_a^{\text{spring}}}{q_s^{\text{spring}}} = \frac{G I_T}{2 E I_r}.
\end{equation}
We are now in a position to quantify the value of $q$ for the two cases below.

\subsection{ Rectangular wire helical  springs }

\begin{figure}
    \centering
    \includegraphics[width=0.6\textwidth]{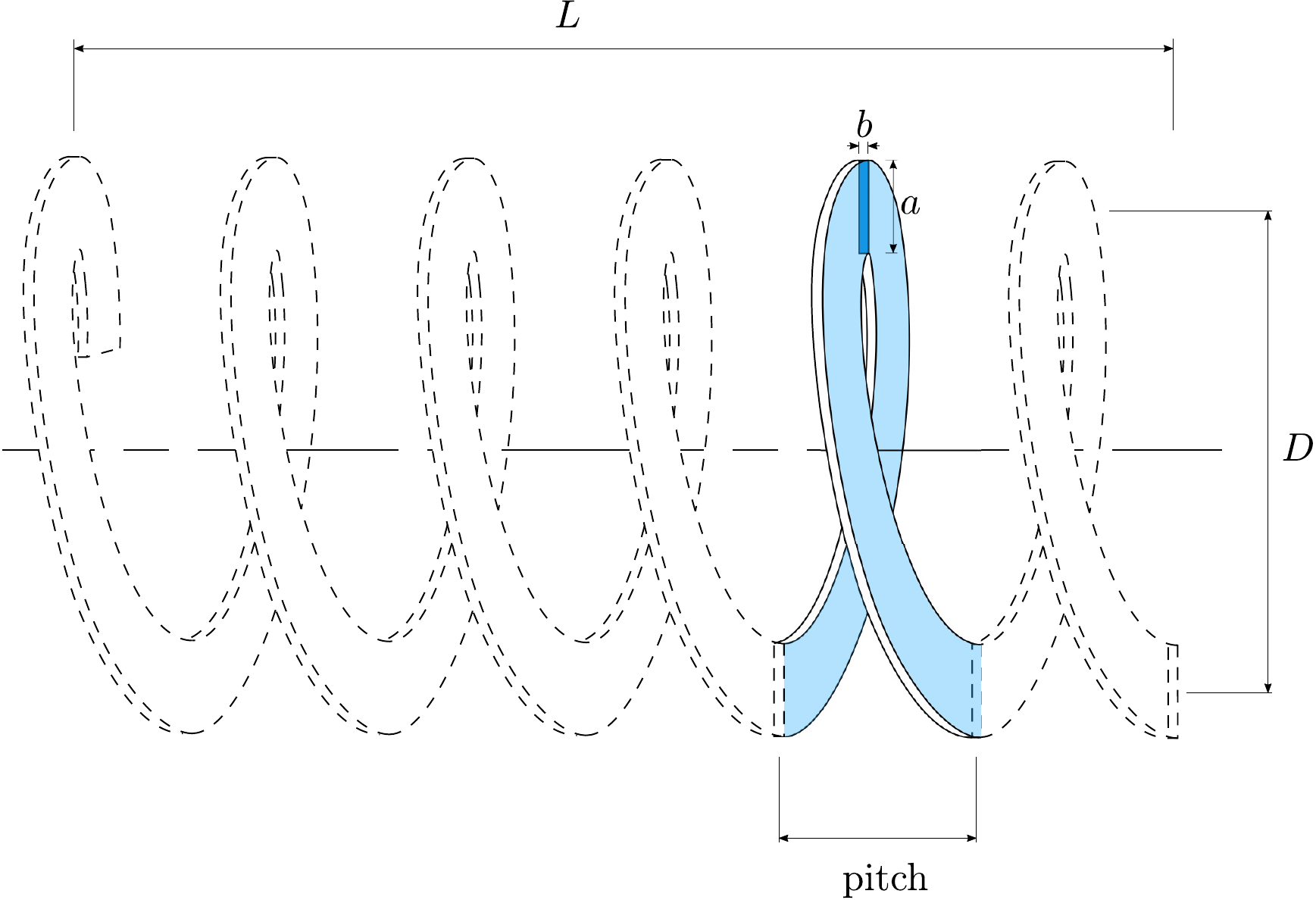}
    \caption{\footnotesize 
        A rectangular wire helical spring realizing an equivalent one-dimensional rod with very small ratios between axial and bending/shear stiffnesses.
        }
    \label{fig:thespring}
\end{figure}

For a rectangular wire of edges $a$ and $b$, as illustrated in Fig. \ref{fig:thespring}, 
the moments of inertia are given by  
\begin{align}
    I = \frac{a b^3}{12} ,\qquad
    I_r = \frac{a^3 b}{12},
\end{align}
and, in the considered case $a>b$ the torsion 
constant by 
\begin{equation}
    I_T\approx \left[1-0.63\frac{b}{a}+0.052\left(\frac{b}{a}\right)^5\right] \frac{a b^3}{3}.
\end{equation}

\begin{figure}
    \centering
    \includegraphics[width=0.8\textwidth]{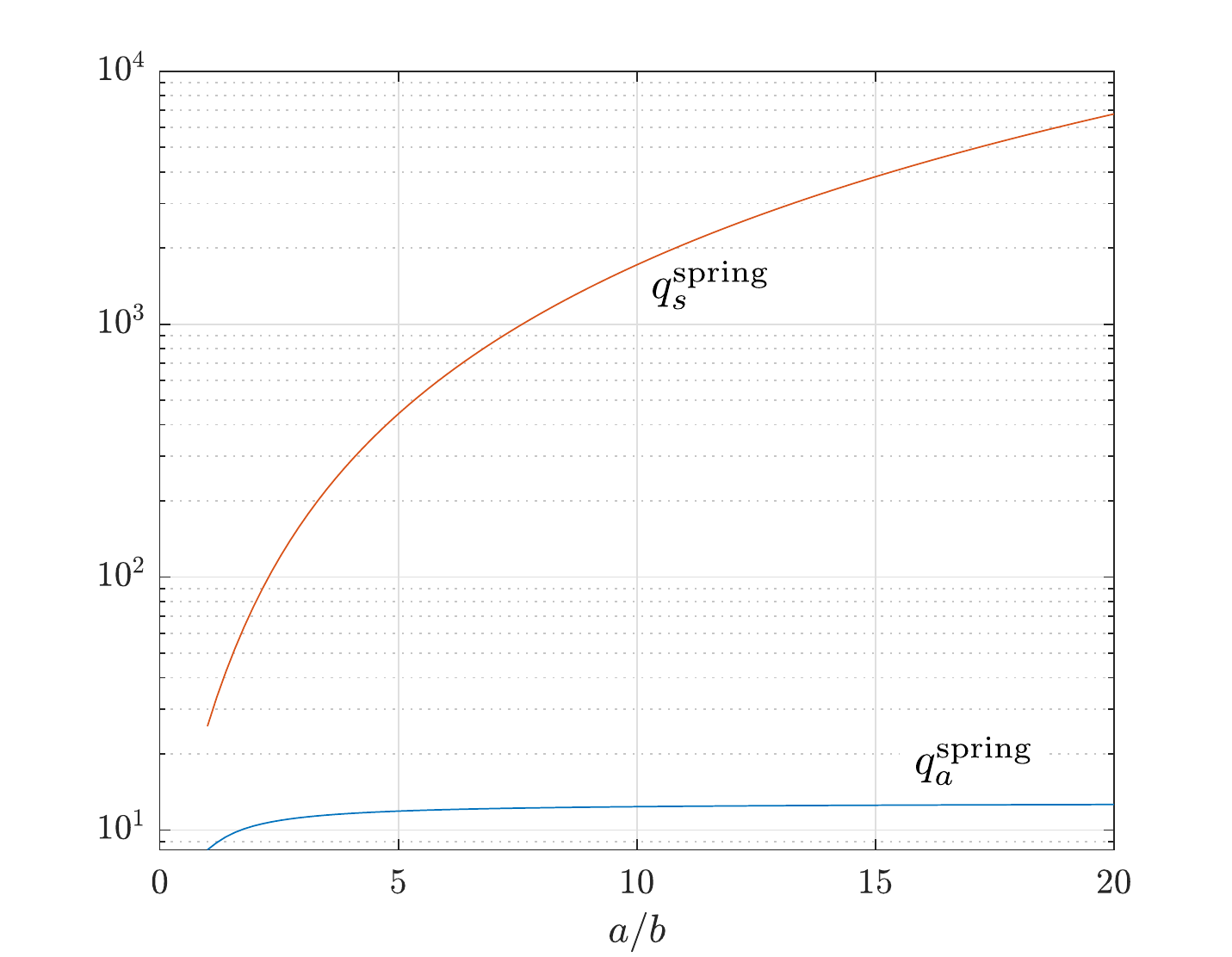}
    \caption{\footnotesize 
    Semi-logarithmic plot of the stiffness ratios $q_a^{\text{spring}}$ and $q_s^{\text{spring}}$ 
    for a rectangular wire helical spring, 
    as functions of the aspect ratio $a/b$ of the rectangular wire cross-section with $L/D=5$ (reported in Fig. \ref{fig:thespring} for $a/b=8$).
        \label{fig:ratios}}
\end{figure}

Considering a steel spring ($E = 2.6$\,$G$, with $G$ being the shear modulus) and characterized by  $L=5D$ (similarly to examples reported  in \cite{Haringx1950OnHC, Haringx1950Ob, wahl1963mechanical}), the stiffness ratios $q_a^{\text{spring}}$ and $q_s^{\text{spring}}$ are shown in Fig. \ref{fig:ratios} as increasing functions of the  aspect ratio $a/b$ of the rectangular cross section. 

Finally, considering for example an aspect ratio $a/b=8$ and  setting the height $b$ of the wire such that the ratio between $b$ and the pitch of the coil is $1/20$, the corresponding spring (shown in Fig. \ref{fig:thespring})  can be compressed up to 95\% of its undeformed length and is characterized by the two following stiffness ratios
\begin{equation}
    q_a^{\text{spring}} = 1.225 \cdot 10,\quad q_s^{\text{spring}} = 1.103 \cdot 10^3,
\end{equation}
providing values for an equivalent rod with negligible shear effects and that could be modelled as the extensible elastica.

\subsection{Composite rod-spring element}

\begin{figure}
    \centering
    \includegraphics[width=0.7\textwidth]{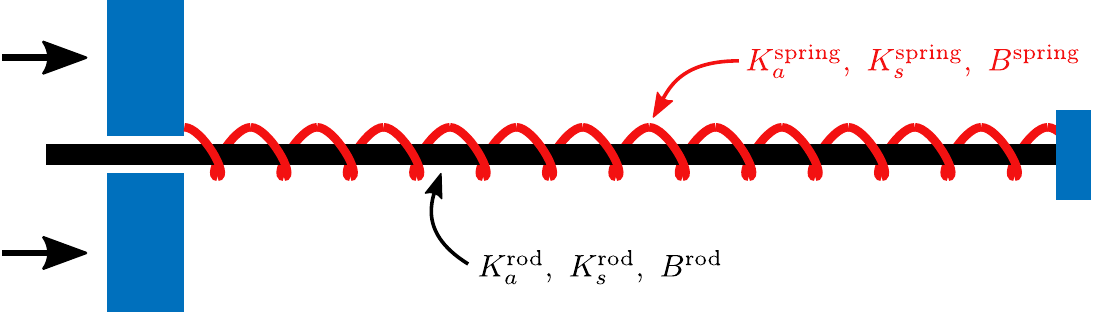}
    \caption{\footnotesize 
      A one-dimensional element working as an axially-deformable elastica, characterized by high shear stiffness ratio $q_s$ but low axial stiffness ratio $q_a$, obtained as  an elastic rod inserted within a helical spring. The spring can slide along the rod so that the latter does not contribute to the axial stiffness of the structure.
    \label{fig:therod}}
\end{figure}

A helical spring is considered to contain a coaxial elastic rod, as shown in Fig. \ref{fig:therod}. While one end of the spring is fixed to the corresponding end of the rod, the remaining part of the spring can  slide without friction along the rod.\footnote{
Although frictionless, the moving boundary problem realized through the relative motion of the spring with respect to the rod may realize configurational forces, whenever a non-null rod's curvature is displayed. Nevertheless, due to their  higher-order character, such configurational forces do not affect the buckling analysis and therefore can be neglected \cite{bigoni2014instability}. This implies that the double restabilization feature is not compromised, so that only the post-buckling response changes.
}
Assuming that  the spring and the rod react  in parallel, the shear and bending  stiffnesses of the equivalent one-dimensional element, working as an axially-deformable elastica,  are simply the sum of that of the spring and of the rod, namely
\begin{align}
     K_{s} = K_{s}^{\text{rod}} +K_{s}^{\text{spring}} , \qquad
 B = B^{\text{rod}} + B^{\text{spring}},
\end{align} 
while, considering that loads are applied on the spring,  the equivalent axial stiffness coincides with that of  the spring because of its frictionless sliding  along the rod,
\begin{align}
     K_{a} =  K_{a}^{\text{spring}}.
\end{align} 
It follows that the axial stiffness can be tuned independently of the bending and shear stiffnesses, to attain (almost) any pair of desired stiffness ratios $q_a$ and $q_s$.

As an example for a round wire spring, the axial, shear and bending stiffnesses are given by 
\begin{equation}
    K_{a}^{\text{spring}} = \frac{G d^4 L}{8 n_a D^3},\quad K_{s}^{\text{spring}} = \frac{E d^4 L}{8 n_a D^3}, \quad B^{\text{spring}} = \frac{E d^4 L}{32 \left(1+\dfrac{E}{2G}\right) n_a D},
\end{equation} 
and for a circular  rod of diameter $D^{\text{rod}}$ as
\begin{equation}
    K_{s}^{\text{rod}} = \frac{8  \pi G \left(D^\text{rod}\right)^2}{37}, \quad B^{\text{rod}} = \frac{\pi E \left(D^\text{rod}\right)^4}{64},
\end{equation}
where for simplicity the rod and the spring have been considered as made up of the same material (differently, the selection of a  material for the rod stiffer than that constituting the spring, $E^{\text{rod}}>E^{\text{spring}}$ further facilitates satisfaction of the inequality $q_a\ll q_s$ with  $q_a\approx 10$). The  stiffness ratios of the equivalent one-dimensional element become 
\begin{equation}\label{panino}
    \begin{aligned}
        & q_{a} = \frac{4 G}{\pi^2 n_a E \beta} \left(\frac{d}{D}\right)^4 \left(\frac{L}{D}\right)^3, \\
        & q_{s} = \frac{1}{\pi^2 \beta} \left(\frac{L}{D}\right)^2 \left[ \frac{4}{n_a} \left(\frac{d}{D}\right)^4 \left(\frac{L}{D}\right)+ \frac{256 \pi G}{37 E} \left(\frac{D^\text{rod}}{D}\right)^2 \right],
    \end{aligned}
\end{equation}
where
\begin{equation}
    \beta = \frac{1}{\left(1+\dfrac{E}{2G}\right) n_a} \left(\frac{d}{D}\right)^4 \left(\frac{L}{D}\right) + \frac{\pi}{2} \left(\frac{D^\text{rod}}{D}\right)^4.
\end{equation}

\begin{figure}
    \centering
    \includegraphics[width=0.8\textwidth]{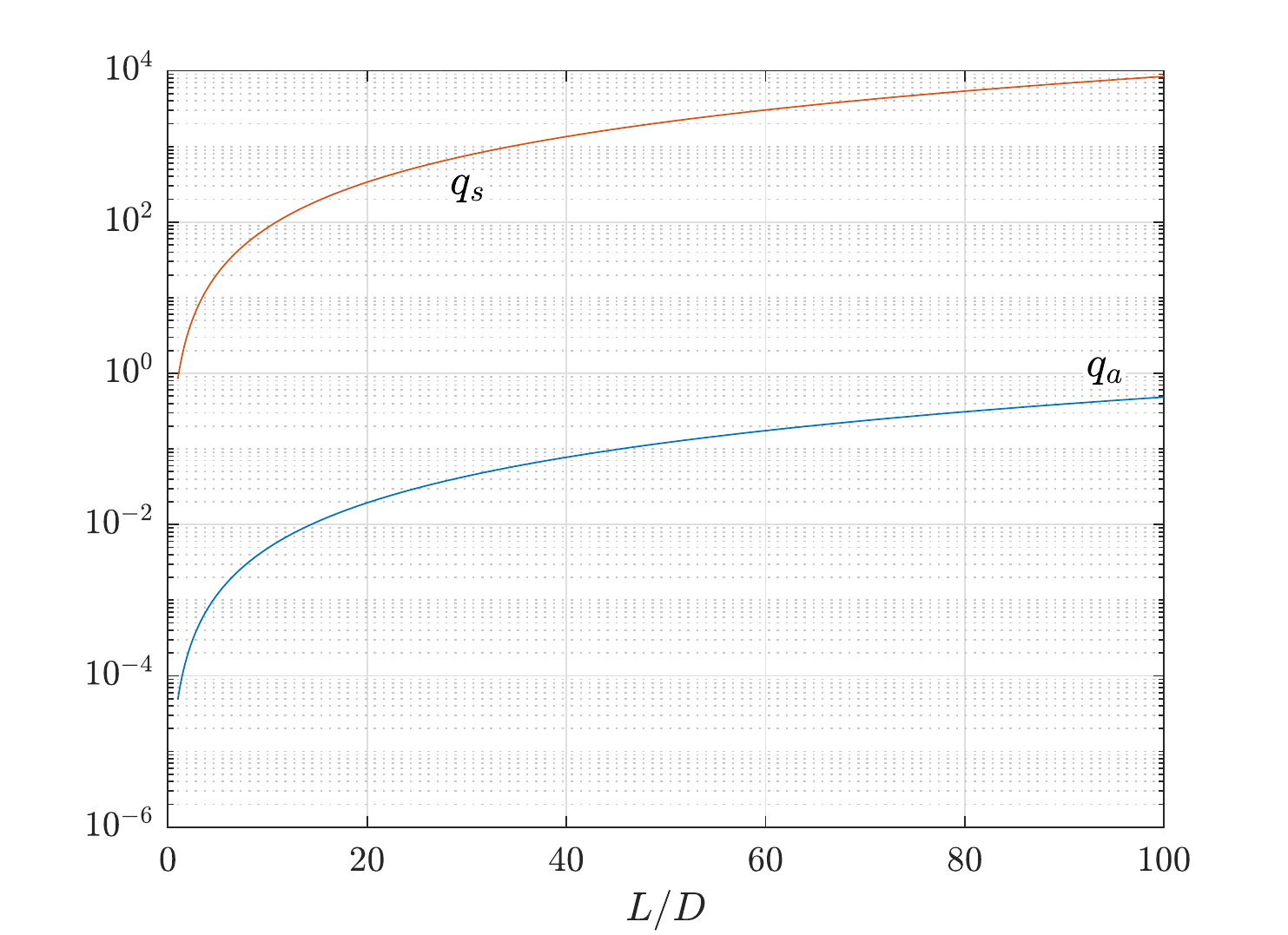}
    \caption{\footnotesize 
    Semi-logarithmic plot of the stiffness ratios $q_a$ and $q_s$ 
    for a composite rod-spring element (Fig. \ref{fig:therod}) 
    as functions of $L/D$.
    \label{fig:ratios_rod}}
\end{figure}

The stiffness ratios described by Eq. (\ref{panino}) are shown in Fig. \ref{fig:ratios_rod} as monotonic increasing functions of the ratio $L/D$, by assuming $d/D=0.1$, $D^\text{rod}/D=0.8$ and the pitch of the coil has been selected in order to have 95\% maximum compression with respect to the undeformed length. Finally, assuming for example $L/D=50$, the stiffness ratios are
\begin{align}
    q_a=1.211\cdot 10^{-1},\qquad q_s=2.107\cdot 10^{3},
\end{align}
corresponding to  values of an equivalent rod with negligible shear effects and that could be properly modelled through the extensible elastica considered in the present paper. 

%%%%%%%%%%%%%%%%%%%%%%%%%%%%%%%%%%%%%%%%%%%%%%%%%%%%%%%%%%%%%%%%%%%%%%%%%%%%%%%%%%%%%%%%%%%%%%%%%%%%%%%%%%%%%%%%%%%%%

\section{Asymptotic behaviour of bifurcation load for vanishing profile curvature at the origin}\label{appendixasymptotic}
\setcounter{equation}{0}
\setcounter{figure}{0} 

\begin{figure}[h]
\centering
    \includegraphics*[width=155mm]{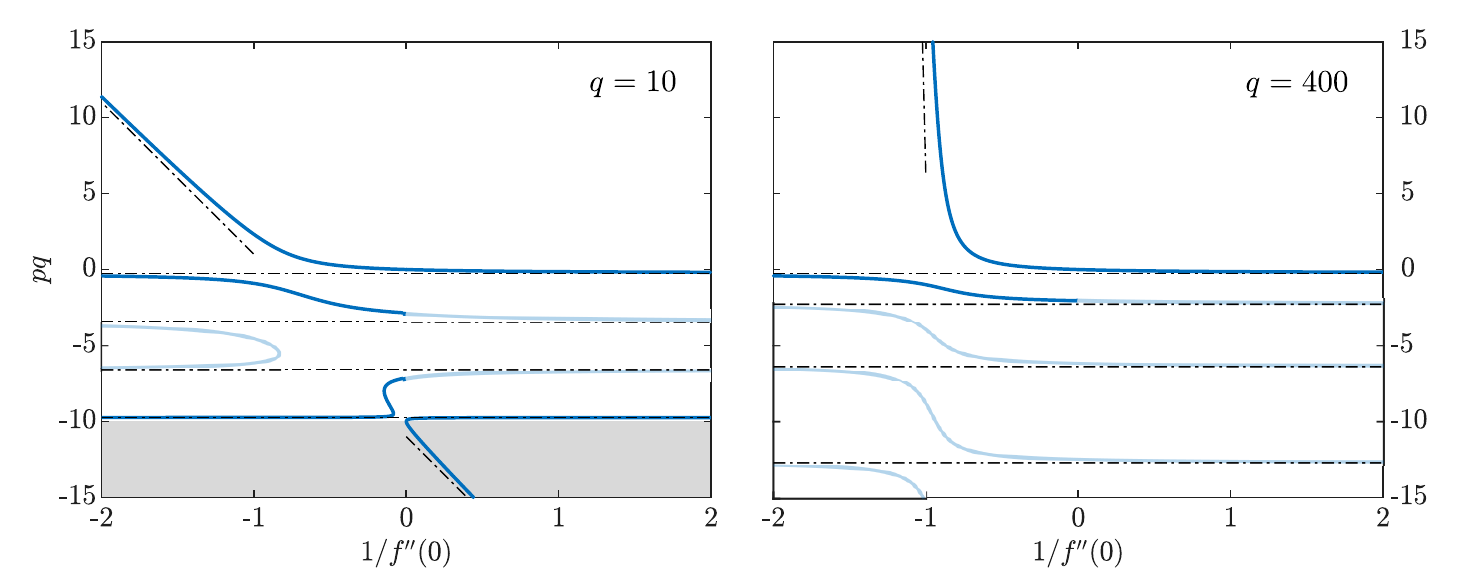}
    \caption{\label{fig:asymptotic} \footnotesize Bifurcation  dimensionless load $p q=P L^2/(\pi^2 B)$ as a function of the dimensionless radius of profile curvature at the origin,  $1/f''(0)$, for fixed values of $q$,   ($q=10$ on the  left  and $q=400$ on the right). The asymptotic behaviour, Eqs.  \eqref{asympttension} and \eqref{asymptcompression}, is represented by  dashed lines.}
\end{figure}

Asymptotic expressions for the bifurcation load in the limit of   vanishing profile curvature at the origin can be obtained by solving the  first-order expansion in $f''(0)\rightarrow0^\pm$ of Eq. \eqref{eq:critical}. It follows that the critical load under tension has the following singular asymptote\footnote{The asymptotic expression \eqref{asympttension} also holds  for  $p$ within the (meaningless) range $(-\infty, -1]\cup[0,\infty)$, as it can be appreciated from Fig. \ref{fig:asymptotic}.}
\begin{equation}
\label{asympttension}
      \lim_{f''(0)\rightarrow 0^-} p_{cr}^{(+)} = -\frac{1}{f''(0)} -1-\frac{\mbox{sgn}(f''(0))}{\pi\sqrt{q}} .
\end{equation}
Concerning the bifurcation loads under compression, $p\in(-1,0)$, the asymptotic behaviour determines the presence of two critical loads (corresponding to destabilization and restabilization of the trivial configuration) for a  number $N$ of critical modes depending on $q$. The asymptotic expression for  the destabilization  $p_{de}^{(-)[n]}$ and restabilization $p_{re}^{(-)[n]}$ loads, pertaining to the $n$-th mode, results to be a finite value independent of the sign of  the profile curvature
\begin{equation}\label{asymptcompression}
    \lim_{f''(0)\rightarrow 0^+}\left\{p_{de}^{(-)[n]},p_{re}^{(-)[n]}\right\} =  \lim_{f''(0)\rightarrow 0^-}\left\{p_{de}^{(-)[n]},p_{re}^{(-)[n]}\right\} = \frac{1}{2} \left( -1 \pm \frac{\sqrt{-1-4 n-4 n^2 + q}}{\sqrt{q}} \right) , 
    \end{equation}
where the number $N$ of existing critical modes is given by
\begin{equation}
    N= \max\left\{\left\lfloor{\dfrac{\sqrt{q}-1}{2}}\right \rfloor,0\right\},
    %\qquad  n (n+1) \leq \frac{q-1}{4},
\end{equation}
with the symbol $\left \lfloor{\cdot}\right \rfloor$ standing for the integer part of the relevant argument. 

It is interesting to note that in the limit of $f''(0)\to0^{\pm}$ the structure reduces to an extensible rod connected to a sliding clamp on its left end and with the right end constrained only in the horizontal direction. In this case, the tensile bifurcation load approaches infinity ($p_{cr}^{(+)}\to\infty$), so that tensile buckling is excluded. 

Finally,  the compressive bifurcation loads in the inextensible limit ($q\to\infty$) match those pertaining to a cantilever Euler beam (with clamped-free boundary conditions), 
\beq
\lim_{q\to\infty}\left\{P_{de}^{(-)[n]}\right\}=-\frac{(2n+1)^2 \pi^2B}{4 L^2},
\eeq
while restabilization does not occur because the corresponding load assumes a negative infinite value
\beq
\lim_{q\to\infty}\left\{P_{re}^{(-)[n]}\right\}=-\infty.
\eeq

As an example, the bifurcation conditions for the dimensionless load $p q$ with varying the dimensionless radius of profile curvature $1/f''(0)$ are reported in Fig. \ref{fig:asymptotic} for $q=10$ (left) and for $q=400$ (right). It can be appreciated that the asymptotic behaviour (drawn as dashed lines) for $1/f''(0)\rightarrow\pm\infty$ is symmetric in the range $p\in[-1,0]$, as predicted by Eq. (\ref{asymptcompression}), and that only one asymptote exists for $1/f''(0)\rightarrow-\infty$ for positive $p$, as predicted by Eq. (\ref{asympttension}).

\end{document}